\documentclass[12pt,a4paper]{article}
\usepackage{amsfonts}
\usepackage{amsmath}
\usepackage{amssymb}
\usepackage{latexsym}

\usepackage[pdftex]{hyperref}

\numberwithin{equation}{section}

\newcommand{\BbbR}{\mathbb{R}}

\DeclareMathOperator{\Realpart}{Re}

\newcommand{\x}{\mathsf{x}}

\title{How often does the Unruh-DeWitt detector click 
beyond four dimensions?}

\author{Lee Hodgkinson\thanks{pmxlh1@nottingham.ac.uk}
\ and
Jorma Louko\thanks{jorma.louko@nottingham.ac.uk}
\\[3ex]
\small{\it School of Mathematical Sciences,
University of Nottingham,}\\
\small{\it Nottingham NG7 2RD, UK}
\\[3ex]
\small{(Revised June 2012)}
\\[3ex]
\small{Published in J.\ Math.\ Phys.\ 
\textbf{53}, 082301 (2012)}
\\ 
{\small\tt \href{http://link.aip.org/link/?jmp/53/082301}{http://link.aip.org/link/?jmp/53/082301}}
}

\date{}

\begin{document}

\maketitle

\begin{abstract}

We analyse the response of an arbitrarily-accelerated
Unruh-DeWitt detector coupled to a massless scalar field in
Minkowski spacetimes of dimensions up to six, working within
first-order perturbation theory and assuming a smooth
switch-on and switch-off. We express the total transition
probability as a manifestly finite and regulator-free
integral formula. In the sharp switching limit, the
transition probability diverges in dimensions greater than
three but the transition rate remains finite up
to dimension five. In dimension six, the 
transition rate remains finite in the sharp switching limit 
for trajectories of constant scalar proper acceleration,
including all stationary trajectories, but it 
diverges for generic trajectories. 
The divergence of 
the transition rate in six dimensions suggests that 
global embedding spacetime (GEMS) methods for 
investigating detector response in curved spacetime 
may have limited validity for generic trajectories 
when the embedding spacetime has dimension higher than five. 
\end{abstract}

\newpage

\section{Introduction}
\label{sec:intro}

The conventional notion of a particle in Minkowski spacetime quantum
field theory is related physically to measurements made by inertial
observers and characterised mathematically by positive frequencies
with respect to the Minkowski time translation Killing vector. This
notion generalises to a certain extent to noninertial observers whose
worldline is the orbit of a more general timelike Killing vector, a
celebrated application being the thermality that uniformly linearly
accelerated observers see in Minkowski vacuum~\cite{unruh}.  The
notion of a particle seen by a generic noninertial observer in
Minkowski space is however significantly more subtle, as a generic
timelike trajectory need not be the orbit of any Killing vector.  The
notion of a particle seen by an observer in curved spacetime is even
more subtle, as a generic curved spacetime need not admit any timelike
Killing vectors~\cite{byd,wald-smallbook}.

The notion of ``a particle seen by an observer" can be given an
operational meaning by coupling the quantum field to a pointlike
particle detector that follows the observer's worldline, known as an
Unruh-DeWitt detector~\cite{unruh,deWitt}, and by regarding
transitions between the detector's quantum states as absorptions and
emissions of particles. This method is applicable whenever the quantum
state is sufficiently regular to render the transition probabilities
well defined, and in particular it does not assume existence of any
Killing vectors. The method confirms the thermality seen by uniformly
linearly accelerated observers in Minkowski vacuum~\cite{unruh}, and
its other applications include the thermality seen by static detectors
in Schwarzschild spacetime \cite{hawking,Hartle:1976tp} and by
inertial detectors in de~Sitter space~\cite{gibb-haw:dS}.

In first-order perturbation theory, the transition probability of the
Unruh-DeWitt detector is proportional to the response function. 
This function is obtained by 
pulling the Wightman distribution of the quantum field 
back to the detector worldline, 
smearing with a switching function that specifies how the 
interaction is turned on and off, and 
Fourier transforming from the detector's proper time to the 
detector's energy level separation. 
The response function is well defined provided the 
quantum state of the field 
is regular in the sense of 
the Hadamard property \cite{kay-wald,Decanini:2005gt}
and the switching function is smooth and has 
compact support~\cite{Fewster:1999gj,junker,hormander-vol1,hormander-paper1}. 
The physical situation here is that the 
initial states 
of the field and the detector have been
prepared before the interaction begins, 
the interaction is switched smoothly on and off, and 
the final state of the detector is read 
after the interaction has ceased. 

It may be tempting to ask what the detector's 
transition probability is while the interaction is still ongoing, 
and to define the transition rate per unit time by differentiating the 
transition probability with respect to the time at which the detector is read. 
The pull-back of the Wightman function is however too singular to 
provide a mathematically well-defined expression for these quantities, and 
seemingly inconspicuous treatments can lead to paradoxical results: 
in particular, a na\"\i{}ve regularisation 
in four-dimensional Minkowski space 
yields a transition rate that is not 
Lorentz covariant~\cite{Sriramkumar:1994pb,schlicht,Schlicht:thesis,louko-satz:profile}. 
In the special cases where the trajectory is stationary, 
the initial state of the field is invariant under the Killing vector
generating the trajectory and the detector has been switched
on in the infinite past
\cite{unruh,deWitt,hawking,gibb-haw:dS}, 
regularisation can be bypassed by
integrating formally 
over the whole trajectory and factoring out the
infinite total proper time, as the transition
rate should be time-independent on grounds of stationarity. 
This line of argument is however not available if the trajectory or 
the initial state of the field are not stationary, as would 
be the case for example for a detector that falls into a black hole, 
or for a detector that responds to the onset of Hawking radiation 
in the spacetime of a collapsing star. 

A mathematically consistent way to formulate the question of detector's response 
while the interaction is still ongoing is to start with a 
smooth switching function of compact support and 
then attempt to take the limit in which the switching function 
approaches a step-function of specified duration. 
In four dimensions, this limit leads to a 
diverging total transition probability within the 
first-order perturbation theory 
treatment~\cite{Sriramkumar:1994pb,Svaiter:1992xt,grove:1988-add,Higuchi:1993cya,Suzuki:1997cz}, 
but the divergent piece depends just on the details of the switching and
neither on the quantum state of the field nor on the detector's trajectory, 
and the derivative of the transition probability with respect 
to the interaction duration remains finite~\cite{satz:smooth,satz-louko:curved}. 
This means that the time derivative of the transition probability provides a 
well-defined notion of a transition rate even in the sharp switching limit. 
Physically, this notion is meaningful when the switching is rapid 
compared with the overall 
duration of the interaction, 
and it then extracts from the 
transition probability the piece that depends on the trajectory and on the quantum state, 
disregarding the numerically dominant piece that only depends on the details of the switching, 
and the whole procedure remains reliable provided the coupling constant 
is so small that first-order perturbation theory is still valid despite 
the large switching-dependent contribution. 
Observationally, the transition rate is meaningful 
in terms of consequent measurements in identical ensembles of detectors~\cite{satz-louko:curved}.
In Minkowski space and with the field prepared 
in Minkowski vacuum, 
this definition of the transition rate agrees with the transition rate 
obtained from the pointlike limit of a spatially smeared but sharply-switched 
detector~\cite{schlicht,Schlicht:thesis,louko-satz:profile,Langlois,Langlois-thesis}. 

In this paper we shall examine the response of an Unruh-DeWitt
detector coupled to a massless scalar field, in first-order
perturbation theory, in the limit of sharp switch-on and switch-off in
Minkowski spacetimes of dimension two to six. Mathematically,
departing from the macroscopically observed spacetime dimension $d=4$
is prompted by the observation that the divergence of the $d=4$
transition probability in the sharp switching limit is due to the
singularity of the Wightman
function~\cite{satz:smooth,satz-louko:curved}, and the strength of
this singularity increases with spacetime dimension. In $d=2$ the
singularity is merely logarithmic in proper time, and the transition
probability remains finite; in $d=4$ the singularity is inverse square
in proper time and the transition probability diverges but the
transition rate remains
finite~\cite{satz:smooth,satz-louko:curved}. Does the divergence of
the total transition probability start already at $d=3$ or only at
$d=4$? Does the instantaneous transition rate become divergent in
sufficiently high dimension, and if yes, is the divergence sensitive
to the non-stationarity of the trajectory?

Physically, one motivation to go above $d=4$ is that our spacetime may
have dimensions not yet observed, or it may admit a holographic
description in terms a higher-dimensional theory.  It is hence of
interest to clarify the usefulness of pointlike particle detectors in
spacetimes that arise in these higher-dimensional theories, including
the various black hole, black brane and black ring spacetimes that
appear in string theory and in the black hole solutions that appear in
brane world cosmology~\cite{bhhdbook-horowitz}.

Another physical motivation to go above $d=4$ is that quantum fields
in curved spacetimes have been modelled by embeddings in a
higher-dimensional flat spacetime, and these global embedding
spacetime (GEMS) methods have yielded reasonable results for the
temperature seen by observers on stationary trajectories in spacetimes
of high symmetry.  A~selection of original papers are
\cite{Deser:1997ri,Deser:1998bb,Deser:1998xb,Santos:2004ws}
and a review with
further references can be found
in~\cite{Langlois,Langlois-thesis}. Could GEMS methods be expected to
provide a reasonable model for the response of an Unruh-DeWitt
detector in the nonstationary setting, for example for a detector
falling into a Schwarzschild black hole, allowing a comparison with
predictions obtained by Bogoliubov transformation
techniques~\cite{Barbado:2011dx}?

As a final motivation, recent investigations of entanglement within
relativistic quantum information theory
\cite{downes-etal,MartinMartinez:2010sg,Dragan:2011zz} have introduced
non-inertial model detectors in a time-dependent setting on an
effectively two-dimensional Minkowski spacetime, where the switch-on
and switch-off can safely be taken sharp. How would the switching
effects need to be handled when the effective spacetime has higher
dimension?

We work in Minkowski spacetime of dimension $d$ and assume the initial
state of the quantum field to be the Minkowski vacuum.  Following the
$d=4$ analysis of \cite{satz:smooth}, we start by taking the switching
function to be smooth and compactly supported, and we express the
response function (proportional to the transition probability by a
factor that only depends on the internal properties of the detector)
as a manifestly regular and Lorentz-covariant integral formula from
which all $i \epsilon$ regulators have been removed. We then
investigate the sharp switching limit, in which the switching function
approaches a theta-step of prescribed duration in a controlled
way. For $d=2$ and $d=3$ the response function remains finite in this
limit; for $d=2$ this is immediate already from the fact that the
singularity in the Wightman function is merely logarithmic.  For $d=4$
it was found in \cite{satz:smooth} that the response function diverges
but its derivative with respect to the duration remains finite, and we
find that the same holds for $d=5$. For $d=6$, the response function
again diverges, but a new phenomenon emerges in its derivative with
respect to the duration: the derivative diverges for precisely those
trajectories on which the scalar proper acceleration is not
constant. This means that for $d=6$ the notion of instantaneous
transition rate is well defined for trajectories of constant scalar
proper acceleration, including all trajectories that are stationary in
the sense of being the orbit of a timelike Killing vector, but is ill
defined for more general trajectories.

As a consistency check, we verify that for uniformly linearly
accelerated motion our instantaneous transition rate results agree in
each dimension with those obtained by Takagi \cite{takagi} by a
regularisation that relies on the stationary at the outset. For $d=6$,
we also verify that our transition rate result agrees with that
obtained by Schlicht's sharply-switched but spatially smeared detector
model~\cite{schlicht,Schlicht:thesis,louko-satz:profile}, in
particular agreeing about the presence of a divergent term that is
proportional to the proper time derivative of the scalar proper
acceleration. This gives us confidence that the obstacle to defining
an instantaneous transition rate for generic trajectories in $d=6$ is
genuine.

Finally, we address the response of a detector in the Schwarzschild
spacetime by the GEMS method. The Kruskal-Szekeres extension of
Schwarzschild \cite{mtw} has a global embedding in six-dimensional
Minkowski space~\cite{fronsdal}, and the GEMS method suggests that
detector response in the Hartle-Hawking-Israel vacuum
\cite{Hartle:1976tp,Israel:1976ur} could be modelled by the response
of the embedded detector in six-dimensional Minkowski space with the
field in Minkowski vacuum. The instantaneous transition rate obtained
from the sharp switching limit of the genuinely four-dimensional
detector is finite~\cite{satz-louko:curved}; by contrast, the GEMS
method predicts a divergent instantaneous transition rate for all
trajectories in which the six-dimensional scalar proper acceleration
is not constant.  Four-dimensional trajectories with constant
six-dimensional scalar proper acceleration include all equatorial
trajectories that have constant area-radius and constant angular
velocity; as special cases, these include the static trajectories and
the circular geodesics. However, we find that the only
four-dimensional timelike \emph{geodesics\/} with constant
six-dimensional scalar proper acceleration are the circular
geodesics. The GEMS method does therefore not provide a way to model
the transition rate of detectors on generic geodesics in
Schwarzschild, including detectors falling freely into the black hole.

We begin in section \ref{sec:detectormodels} by reviewing the
Unruh-DeWitt detector model, emphasising how the regularisation of the
Wightman distribution needs to be addressed prior to the sharp
switching limit.  In section \ref{sec:6d} we write the $d=6$ response
function in a form in which the regulator in the Wightman distribution
has been eliminated, adapting the $d=4$ techniques
of~\cite{satz:smooth}.  In sections \ref{sec:3d} and \ref{sec:5d} we
perform a similar analysis for respectively $d=3$ and $d=5$, where new
technical issues arise from the fractional powers of squared geodesic
distance that occur in odd spacetime dimensions.  The sharp switching
limit is addressed in section~\ref{sec:sharp}.  Section
\ref{sec:consistencychecks} presents an overview of the divergences
that emerge as the spacetime dimension increases, giving consistency
checks with earlier work and discussing how the structure might
continue for $d>6$.  Section \ref{sec:GEMS} applies the results to
extended Schwarzschild as embedded in $d=6$ Minkowski.  Section
\ref{sec:summary} presents a summary and concluding remarks.

We work in Minkowski spacetime of dimension 
$d \ge2$ and of metric signature $({-} {+}\cdots{+})$. 
The units are such that $\hbar= c=1$. 
Sans-serif letters denote Minkowski $d$-vectors, 
the Minkowski scalar product of $d$-vectors 
$\mathsf{k}$ and $\mathsf{x}$
is denoted by 
$\mathsf{k}\cdot \mathsf{x}$, 
and we write $\mathsf{x}^2 := \mathsf{x}\cdot \mathsf{x}$. 
Complex conjugation is denote by overline. 
$O(x)$ denotes a quantity for which 
$O(x)/x$ is bounded as $x\to0$, 
and $O(1)$ denotes 
a quantity that remains bounded when the 
parameter under consideration approaches zero.

\section{Particle detector model}
\label{sec:detectormodels}

We take as our detector model a point-like two-state `atom'. The
detector Hilbert space is spanned by the orthonormal basis states
$|0\rangle_d$ and $|E\rangle_d$ whose respective energy eigenvalues
are $0$ and~$E$, with $E\ne0$. For $E>0$, $|0\rangle_d$ is the ground
state and $|E\rangle_d$ is the excited state; for $E<0$, the roles are
reversed.

The detector moves in $d$-dimensional Minkowski space along the 
timelike worldline~$\x(\tau)$, where the parameter $\tau$ is the detector's proper time. 
The detector is coupled to a real, massless scalar field $\phi$ by the interaction Hamiltonian 
\begin{equation}
H_{\text{int}}=c\chi(\tau)\mu(\tau)\phi\bigl(\x(\tau)\bigr) 
\ , 
\end{equation}
where $c$ is a small coupling constant, $\mu(\tau)$ is 
the atom's monopole moment operator and $\chi(\tau)$ 
is a switching function, positive during the 
interaction and vanishing elsewhere.

We assume the switching function $\chi(\tau)$ to be smooth 
and of compact support, and we assume the trajectory 
$\x(\tau)$ to be smooth on the support of~$\chi$. 
We shall return to the 
smoothness of the trajectory in 
section~\ref{sec:summary}. 

We take the detector to be initially in the state $|0\rangle_d$ and the field to be in the Minkowski vacuum $|0\rangle$. After the interaction has been turned on and off, we are interested in the probability for the detector to have made a transition to the state~$|E\rangle_d$, regardless the final state of the field. In first order perturbation theory in~$c$, this probability factorises as~\cite{byd,wald-smallbook}
\begin{equation}
\label{eq:prob}
P(E)=c^2{|_d\langle0|\mu(0)|E\rangle_d|}^2\mathcal{F}\left(E\right)
\ , 
\end{equation}
where the factor ${|_d\langle0|\mu(0)|E\rangle_d|}^2$ depends only on the internal structure of the detector but neither on the trajectory nor on the switching, while the dependence on the trajectory and on the switching is encoded in the response function~$\mathcal{F}\left(E\right)$. The formula for the response function reads 
\begin{equation}
\label{eq:respfunc-def}
\mathcal{F}\left(E\right)=\int^{\infty}_{-\infty}\,\mathrm{d}\tau'\,\int^{\infty}_{-\infty}\,\mathrm{d}\tau''\, e^{-iE(\tau'-\tau'')} \,\chi(\tau')\chi(\tau'') \, W(\tau',\tau'')
\ , 
\end{equation}
where the correlation function 
$W(\tau',\tau'') :=\langle0|\phi\bigl(\x(\tau')\bigr)\phi\bigl(\x(\tau'')\bigr)|0\rangle$ is the pull-back of the Wightman function to the detector world line. Using $\overline{W(\tau',\tau'')} = W(\tau'',\tau')$ and changing the integration variables from $(\tau',\tau'')$ to $(u,s)$ where $u := \tau'$ and $s:=\tau''-\tau'$, a useful alternative expression for the response function is~\cite{schlicht}
\begin{equation}
\label{eq:respfunc-alt}
\mathcal{F}\left(E\right)=2 \Realpart 
\int^{\infty}_{-\infty}\,\mathrm{d}u\,\chi(u)\int^{\infty}_0\,\mathrm{d}s\,\chi(u-s)\,e^{-iEs} \, W(u,u-s)
\ . 
\end{equation}

In summary, we may regard the response function as answering the question 
``What is the probability of the detector 
to be observed in the state $|E\rangle_d$ after the 
interaction has ceased?" With mild abuse of terminology, we 
shall drop the constant prefactors in \eqref{eq:prob} 
and refer to the response function simply as the transition probability. 

As the detector's worldline is by assumption timelike, 
it follows from the general properties of the wave front 
sets of Hadamard state Wightman functions that 
the correlation function $W$ is a well-defined distribution on 
$\BbbR\times\BbbR$~\cite{Fewster:1999gj,junker,hormander-vol1,hormander-paper1}. 
As the switching function is by assumption smooth with compact support, formulas \eqref{eq:respfunc-def} and \eqref{eq:respfunc-alt} are hence mathematically well defined. They are however not as such well suited for discussing the sharp switching limit. While the distribution $W$ may be represented by a family $W_\epsilon$ of functions that converge to $W$ as $\epsilon\to0_+$, the limit $\epsilon\to0_+$ may not necessarily be taken in \eqref{eq:respfunc-def} and \eqref{eq:respfunc-alt} pointwise under the integrals. Our first task is therefore to express the response function in a form in which the regulator $\epsilon$ does not appear. 

The case $d=2$ is exceptional. The Wightman function of a massless scalar field in two dimensions is infrared divergent and should be understood in some appropriate limiting sense, such as the $m\to0$ limit of a scalar field of mass~$m>0$. Given this understanding, the singularity in the correlation function $W(\tau',\tau'')$ is logarithmic in $\tau'-\tau''$, that is, integrable. In this case it follows by dominated convergence that the sharp switching limit in \eqref{eq:respfunc-alt} can be taken immediately by setting $\chi(u)=\theta(u-\tau_0)\theta(\tau-u)$, where $\theta$ is the Heaviside function, $\tau_0$~is the moment of switch-on and $\tau$ is the moment of switch-off. The result is 
\begin{equation}
\label{eq:tottranprob}
\mathcal{F}_{\tau}\left(E\right)= 
2 \Realpart 
\int^{\tau}_{\tau_0}\,\mathrm{d}u\,
\int^{u-\tau_0}_0\,\mathrm{d}s\,\,e^{-iEs} \, W(u,u-s) 
\ . 
\end{equation}
The instantaneous transition rate can then be defined as the derivative of $\mathcal{F}_{\tau}\left(E\right)$ (\ref{eq:tottranprob}) with respect to $\tau$, with the result \cite{Langlois-thesis,white-dissertation} 
\begin{equation}
\label{eq:tranrate}
\dot{\mathcal{F}}_{\tau}\left(E\right)=2 \Realpart 
\int^{\Delta\tau}_0\,\mathrm{d}s\,\,e^{-iEs} \, W(\tau,\tau-s)
\ , 
\end{equation}
where $\Delta\tau:=\tau-\tau_0$. 

For $d>2$, the singularity in $W(\tau',\tau'')$ is proportional to ${(\tau'-\tau'')}^{2-d}$, and the regulator must be removed more carefully. The case $d=4$ was addressed in~\cite{satz:smooth}. In the following sections we shall address the cases $d=6$, $d=3$ and $d=5$ in turn.

\section{Response function for $d=6$}
\label{sec:6d}

In this section we remove the regulator from the response function formula \eqref{eq:respfunc-alt} for $d=6$. The method is an adaptation of that introduced in \cite{satz:smooth} for $d=4$. 

\subsection{Regularisation}
\label{sec:6d:reg}

The regularised $d=6$ Wightman function 
reads \cite{kay-wald,Decanini:2005gt,Langlois-thesis}
\begin{equation}
\label{eq:w6d}
W_\epsilon(u,u-s)=\frac{1}{4\pi^3}
\frac{1}{\left[{(\Delta\x)}^2+2i\epsilon\Delta t+\epsilon^2\right]^2}
\ , 
\end{equation}
where $\epsilon>0$ is the regulator, 
$\Delta\x:=\x(u)-\x(u-s)$ and $\Delta t:=t(u)-t(u-s)$. 
From \eqref{eq:respfunc-alt} we obtain 
\begin{equation}
\begin{aligned}
\label{eq:detres6d}
\mathcal{F}(E)&=\lim_{\epsilon\to 0}
\frac{1}{2\pi^3} 
\int^{\infty}_{-\infty}\,\mathrm{d}u\,\chi (u)\int^{\infty}_{0}\,
\mathrm{d}s\,
\frac{\chi(u-s)}{R^4} \, \times 
\\[1ex]
&\times 
\left[
\cos{(Es)}\bigl[\bigl((\Delta\x)^2+\epsilon^2\bigr)^2-4\epsilon^2{(\Delta t)}^2\bigr] 
-4\epsilon\sin{(Es)}\Delta t \bigl({(\Delta\x)}^2+\epsilon^2\bigr) 
\right]  
\end{aligned}
\end{equation}
with 
\begin{align}
\label{eq:R-def}
R := 
\sqrt{\left[{(\Delta\x)}^2+\epsilon^2\right]^2+4\epsilon^2{(\Delta t)}^2}
\ , 
\end{align}
where in \eqref{eq:R-def} the quantity under the square root is
positive and the square root is taken positive. 

We record here  
inequalities that will be used repeatedly below. First, as
geodesics maximise the proper time on 
timelike curves in Minkowski space, it follows 
that $|(\Delta \x)^2| \geq s^2$. Second, as 
$\chi$ has compact support,  
the contributing interval of $s$ in \eqref{eq:detres6d} is bounded above, 
uniformly under the integral over~$u$. From 
the small $s$ expansions $(\Delta\x)^2=-s^2+O\left(s^4\right)$
and $\Delta t = O\left(s\right)$
it hence follows that
$|(\Delta\x)^2| \leq Ks^2$ 
and 
$|\Delta t| \leq sM$, where $K$ and $M$ 
are positive constants, independent of~$u$.

We need to address first the integral over $s$ in~\eqref{eq:detres6d}. 
Working under the expression 
${(2\pi^3)}^{-1} \int^{\infty}_{-\infty}\,\mathrm{d}u\,\chi (u)$, 
we write this integral over $s$ as the sum 
$I^{\text{even}}_{<} + I^{\text{odd}}_{<} + I^{\text{even}}_{>} + I^{\text{odd}}_{>}$, 
where the superscript even (odd) refers to the factor 
$\cos(Es)$ (respectively $\sin(Es)$) and the subscript $<$ ($>$) 
indicates that the range of integration is 
$(0,\eta)$ (respectively $(\eta,\infty)$), where 
$\eta:=\epsilon^{1/4}$. 
We remark that this choice for 
$\eta$ differs from the choice 
$\eta=\epsilon^{1/2}$ that was 
made for $d=4$ in \cite{louko-satz:profile,satz:smooth} 
and will be made for $d=3$
in section \ref{sec:3d} below, 
for reasons that stem from the increasing singularity of the 
Wightman function with increasing~$d$. 

We consider the two intervals of $s$ in the next two subsections. 

\subsection{Subinterval $\eta < s < \infty$}
\label{sec:6d:larges}

Consider $I^{\text{even}}_{>}$. When 
$\epsilon$ is set to zero, the integrand in $I^{\text{even}}_{>}$
reduces to $\chi(u-s)\cos{(Es)}/\left[(\Delta\x)^2\right]^2$. 
This replacement creates in $I^{\text{even}}_{>}$ 
an error that can be arranged in the form 
\begin{equation}
\begin{aligned}
\label{eq:error6d}
\int^{\infty}_{\eta}\,\mathrm{d}s\, &\chi(u-s) \, 
\frac{\cos{(Es)}}{\left[(\Delta\x)^2\right]^2} \, \times\\
& \times 
\left[\frac{\left(\left(1+\frac{\epsilon^2}{(\Delta\x)^2}\right)^2-4\epsilon^2\frac{(\Delta t)^2}{\left[(\Delta\x)^2\right]^2} \right)-\left(\left(1+\frac{\epsilon^2}{(\Delta\x)^2}\right)^2+4\epsilon^2\frac{(\Delta t)^2}{\left[(\Delta\x)^2\right]^2} \right)^2}{\left(\left(1+\frac{\epsilon^2}{(\Delta\x)^2}\right)^2+4\epsilon^2\frac{(\Delta t)^2}{\left[(\Delta\x)^2\right]^2} \right)^2}\right] .
\end{aligned}
\end{equation}
Using $|(\Delta \x)^2| \geq s^2$ and $s\ge \eta = \epsilon^{1/4}$, we have 
$|\epsilon^2/(\Delta\x)^2 |\leq \epsilon^2/s^2 \leq \epsilon^2/\sqrt{\epsilon} =O\left(\epsilon^{3/2}\right)$. Using $|\Delta t| \leq sM$, we similarly have 
$\epsilon^2 (\Delta t)^2/\left[(\Delta\x)^2\right]^2 =O\left(\epsilon^{3/2}\right)$.
The integrand in 
\eqref{eq:error6d} is hence bounded in absolute value by 
a constant times $\epsilon^{3/2}/\left[(\Delta\x)^2\right]^2 \le \epsilon^{3/2}/s^4$. 
It follows that the integral is of order $O\left(\epsilon^{3/2}/\eta^3\right) = O\left(\eta^3\right)$. 

Similar estimates show that $I^{\text{odd}}_{>}=O\left(\eta\right)$. 

Collecting, we have 
\begin{equation}
\label{eq:6d:upper}
I^{\text{even}}_{>}+I^{\text{odd}}_{>}
=\int^{\infty}_{\eta}\,\mathrm{d}s\,
\frac{\chi(u-s)\cos{(Es)}}{\left[(\Delta\x)^2\right]^2}+O\left(\eta\right) 
\ . 
\end{equation}

\subsection{Subinterval $0 < s < \eta$}
\label{sec:6d:smalls}

Consider $I^{\text{odd}}_{<}$, given by 
\begin{equation}
\label{eq:6dlessodd}
I^{\text{odd}}_{<}=-4\epsilon\int^{\eta}_{0}\,\mathrm{d}s\,\chi(u-s)
\, 
\frac{\sin{(Es)}\Delta t\bigl({(\Delta\x)}^2+\epsilon^2\bigr)}
{R^4}
\ . 
\end{equation}
The delicate task is to estimate the 
denominator in~\eqref{eq:6dlessodd}. 

By Taylor's theorem, 
$(\Delta t)^2$, $(\Delta\x)^2$ and $\left[(\Delta\x)^2\right]^2$ have the asymptotic 
small $s$ expansions 
\begin{subequations}
\label{eq:tDeltaandSq-expansions}
\begin{align}
(\Delta t)^2 & =\sum^{n_1-1}_{n=0} T_n s^n 
+ O(s^{n_1})
\ , 
\\
(\Delta\x)^2 & =\sum^{n_2-1}_{n=0} X_n s^n 
+ O(s^{n_2})
\ , 
\label{eq:Deltaxexp}
\\
\left[(\Delta\x)^2\right]^2 & =\sum^{n_3-1}_{n=0} F_n s^n 
+ O(s^{n_3})
\ , 
\end{align}
\end{subequations}
where the expansion coefficients 
$T_n$, 
$X_n$ and 
$F_n$ are functions of~$u$, satisfying 
$T_0 = T_1 = X_0 = X_1 = X_3=0$, $X_2 = -1$ 
and the consequences for~$F_n$, 
and the positive integers $n_1$, $n_2$ and $n_3$ may be chosen arbitrarily. 
(Note that as the trajectory is assumed smooth but not necessarily analytic, 
the error terms in \eqref{eq:tDeltaandSq-expansions} 
are not guaranteed to vanish for fixed $s$ as $n_i\to\infty$.)  
With this notation, the key insight is the rearrangement 
\begin{align}
\label{eq:den6d-raw}
R^2 
& = 
\epsilon^{4}P\Bigg[1-\frac{4\dot{t}\ddot{t}\epsilon r^3}{P}
+\sum^{n_4-1}_{n=0}\left(2X_{(n+4)}+4T_{(n+4)}\right)\frac{\epsilon^{n+2}r^{n+4}}{P}
+\sum^{n_5-1}_{n=0}F_{(n+6)}\frac{\epsilon^{n+2}r^{n+6}}{P}
\nonumber 
\\[1ex]
& \hspace{10ex}
+ \frac{O\bigl(\epsilon^{n_4+2}r^{n_4+4}\bigr)}{P}
+ \frac{O\bigl(\epsilon^{n_5+2}r^{n_5+6}\bigr)}{P}
\Biggr]
\ , 
\end{align}
where we have written 
$s = \epsilon r$,  
\begin{align}
P :=  1+2(2\dot{t}^2-1)r^2+r^4  
\ , 
\end{align} 
the dots denote derivatives with respect to~$u$, 
and the positive integers $n_4$ and $n_5$ may be chosen arbitrarily. 

Since $\dot{t}\geq 1$, $P$ is positive for $r \ge0$. 
We wish to regard the external factor $\epsilon^4 P$ 
in \eqref{eq:den6d-raw} as the dominant part and the 
terms in the square brackets as a leading 1 plus 
sub-leading corrections. 
To this end, we rewrite \eqref{eq:den6d-raw} as 
\begin{align}
R^2 
&= 
\epsilon^{4}P\Bigg[1-\frac{4\dot{t}\ddot{t}\epsilon r^3}{P}z
+\sum^{n_4-1}_{n=0}\left(2X_{(n+4)}+4T_{(n+4)}\right)\frac{\epsilon^{n+2}r^{n+4}}{P}z^{2+(n/4)}
\nonumber
\\[1ex]
& 
+\sum^{n_5-1}_{n=0}F_{(n+6)}\frac{\epsilon^{n+2}r^{n+6}}{P}z^{(n+2)/4}
+ \frac{O\bigl(\epsilon^{n_4+2}r^{n_4+4}\bigr)}{P}z^{2+(n_4/4)}
+ \frac{O\bigl(\epsilon^{n_5+2}r^{n_5+6}\bigr)}{P}z^{(n_5+2)/4}
\Bigg] \ , 
\label{eq:den6d}
\end{align}
where the book-keeping parameter $z$, with numerical value~$1$, 
indicates what order in $\epsilon$ the term 
in question is \emph{uniformly\/} over the full range of~$r$, 
$0\le r \le \epsilon^{-3/4}$.  
The term 
$-4\dot{t}\ddot{t}\epsilon r^3/P$
is assigned the factor $z$ because 
$r^3/P$ is bounded by a constant. 
The $z$-factors in the other terms follow because 
$r^{n+4}/P$ and $r^{n+6}/P$ are respectively 
bounded by 
a constant times 
$\epsilon^{-3n/4}$ and a constant times $\epsilon^{-3(n+2)/4}$. 

Inserting \eqref{eq:den6d} in the denominator of \eqref{eq:6dlessodd},
performing a similar expansion in the numerator and changing the
integration variable to $r$ we obtain
\begin{equation}
\begin{aligned}
\label{eq:6d:odd:premult}
& 
I^{\text{odd}}_{<} = -\frac{1}{\epsilon^2}\int^{\eta^{-3}}_{0}\,\mathrm{d}r\,\frac{4Er^2}{P} \left[\chi(u)-\epsilon r\dot{\chi}(u)z^{1/4}+\cdots+\frac{1}{8!}\chi^{(8)}(u)\epsilon^8 r^8 z^2+O\left(z^{9/4}\right)\right]\times \\
&\ \times \left[1-\frac{1}{3!}\epsilon^2E^2r^2\sqrt{z}+\cdots+\frac{1}{9!}\epsilon^8 E^8 r^8 z^2+O\left(z^{5/2}\right)\right]\times \\
&\ \times \left[\dot{t}-\frac{1}{2}\ddot{t}\epsilon r z^{1/4}+\cdots+\frac{1}{9!}t^{(9)}\epsilon^8 r^8 z^2+O\left(z^{9/4}\right)\right]\left[\frac{1-r^2}{P}+X_4\frac{\epsilon^2 r^4}{P}z^2+O\left(z^{9/4}\right) \right] \times\\
&\ \times \Bigg[1-\frac{4\dot{t}\ddot{t}\epsilon r^3}{P}z+\left(2X_4+4T_4\right)\frac{\epsilon^{2}r^{4}}{P}z^{2}+\sum^{6}_{n=0}F_{(n+6)}\frac{\epsilon^{n+2}r^{n+6}}{P}z^{(n+2)/4}+O\left(z^{9/4}\right)\Bigg]^{-2} , 
\end{aligned}
\end{equation}
where in each bracket factor in the integrand we have kept terms to order $z^2$ because 
of the factor $\epsilon^{-2}$ outside the integral and because 
$\int^{\eta^{-3}}_{0} (r^2/P)\, \mathrm{d}r$ remains bounded as $\eta\to0$. 
We may now Taylor expand the integrand 
in \eqref{eq:6d:odd:premult} in~$z^{1/4}$, 
keeping terms to order~$z^2$: 
the dropped terms are of order $z^{9/4}$ and their contribution to 
$I^{\text{odd}}_{<}$ is~$O(\eta)$.
After this expansion $z$ can be replaced by its numerical value~$1$, and we obtain
for $I^{\text{odd}}_{<}$ an expression that 
consists of elementary integrals of rational functions 
plus the error term~$O(\eta)$.

Consider then $I^{\text{even}}_{<}$, 
rearranged as 
\begin{equation}
\begin{aligned}
\label{eq:6dlesseven}
I^{\text{even}}_{<}
&=
\int^{\eta}_{0}\,\mathrm{d}s\,
\chi(u-s) \, \frac{\cos{(Es)}}{R^2}
\\[1ex]
&
\hspace*{3ex}
-8\epsilon^2\int^{\eta}_{0}\,\mathrm{d}s\,
\chi(u-s) \, \frac{\cos{(Es)} {(\Delta t)}^2}{R^4}
\ . 
\end{aligned}
\end{equation}
Proceeding as above, we find 
\begin{align}
&I^{\text{even}}_{<}
=\frac{1}{\epsilon^3}\int^{\eta^{-3}}_{0}\,\frac{\mathrm{d}r}{P}\,\left[\chi(u)-\epsilon r\dot{\chi}(u)z^{1/4}+\cdots+\frac{1}{12!}\chi^{(12)}(u)\epsilon^{12} r^{12} z^3+O\left(z^{13/4}\right)\right]\times 
\notag 
\\
&\hspace{17ex}
\times \left[1-\frac{1}{2!}\epsilon^2E^2r^2\sqrt{z}+\cdots+\frac{1}{12!}\epsilon^{12} E^{12} r^{12} z^3+O\left(z^{7/2}\right)\right]\times 
\notag 
\\
&\hspace{17ex}
\times \Bigg[1-\frac{4\dot{t}\ddot{t}\epsilon r^3}{P}z+\sum^{4}_{n=0}\left(2X_{(n+4)}+4T_{(n+4)}\right)\frac{\epsilon^{(n+2)}r^{(n+4)}}{P}z^{2+(n/4)}
\notag 
\\
&\hspace{23ex}
+\sum^{10}_{n=0}F_{(n+6)}\frac{\epsilon^{n+2}r^{n+6}}{P}z^{(n+2)/4}+O\left(z^{13/4}\right)\Bigg]^{-1} 
\notag 
\\
&\hspace{8ex}
-\frac{8}{\epsilon^3}\int^{\eta^{-3}}_{0}\,\mathrm{d}r\,\left[\chi(u)-\epsilon r\dot{\chi}(u)z^{1/4}+\cdots-\frac{1}{15!}\chi^{(15)}(u)\epsilon^{15} r^{15} z^{15/4}+O\left(z^{4}\right)\right]\times 
\notag 
\\
&\hspace{20ex}
\times \left[1-\frac{1}{2!}\epsilon^2E^2r^2\sqrt{z}+\cdots-\frac{1}{14!}\epsilon^{14} E^{14} r^{14} z^{7/2}+O\left(z^{4}\right)\right]\times 
\notag 
\\
&\hspace{20ex}
\times \left[\frac{T_2r^2}{P^2}+\frac{T_3\epsilon r^3}{P^2}z+\frac{T_4\epsilon^2 r^4}{P^2}z^2+\frac{T_5\epsilon^3 r^5}{P^2}z^3+O\left(z^{4}\right)\right]\times 
\notag 
\\
&\hspace{20ex}
\times \Bigg[1-\frac{4\dot{t}\ddot{t}\epsilon r^3}{P}z+\sum^{7}_{n=0}\left(2X_{(n+4)}+4T_{(n+4)}\right)\frac{\epsilon^{(n+2)}r^{(n+4)}}{P}z^{2+(n/4)}
\notag 
\\
&\hspace{25.5ex}
+\sum^{13}_{n=0}F_{(n+6)}\frac{\epsilon^{n+2}r^{n+6}}{P}z^{(n+2)/4}+O\left(z^{4}\right)\Bigg]^{-2}
\ . 
\label{eq:6d:even:premult}
\end{align}
We Taylor expand the integrands in \eqref{eq:6d:even:premult}
in~$z^{1/4}$, 
keeping in the first (respectively second) integrand terms to order 
$z^3$~($z^{15/4}$), 
at the expense of an error of order $O(\eta)$ in~$I^{\text{even}}_{<}$. 
Replacing $z$ 
by its numerical value $1$, we then obtain 
for $I^{\text{even}}_{<}$ an expression that 
consists of elementary integrals of rational functions 
plus the error term~$O(\eta)$.

\subsection{Combining the subintervals}

Evaluating the numerous elementary integrals 
obtained from \eqref{eq:6d:odd:premult} and 
\eqref{eq:6d:even:premult}
and combining the results with~\eqref{eq:6d:upper}, 
we find from \eqref{eq:detres6d} that the response function takes the form 
\begin{equation}
\begin{aligned}
\label{eq:6d:result}
\mathcal{F}(E)&=\lim_{\eta\to 0}\frac{1}{2\pi^3} \int^{\infty}_{-\infty}\,\mathrm{d}u\,\chi (u)\Bigg[ -\frac{\chi(u)}{3\eta^3}-\frac{E\pi}{12}\left[\chi(u)(E^2+\ddot{\x}^2)-3\ddot{\chi}(u)\right]\\
&+\frac{1}{6\eta}\left[\chi(u)(3E^2+\ddot{\x}^2)-3\ddot{\chi}(u)\right] +\int^{\infty}_{\eta}\,\mathrm{d}s\,\frac{\chi(u-s)\cos{(Es)}}{\left[(\Delta\x)^2\right]^2}\Bigg]
\ , 
\end{aligned}
\end{equation}
where $\ddot{\x}^2$ is evaluated at~$u$. 
The uniformity of the $O(\eta)$ error terms in 
$u$ has been used to control the errors, and 
all terms involving the Lorentz-noncovariant quantities 
$T_n$ have cancelled on integration 
over $u$ (cf.~section 3 of \cite{satz:smooth} 
for a similar cancellation in four dimensions). 
Taking the inverse powers of $\eta$ 
under the $s$-integral, we have 
\begin{equation}
\label{eq:6d:result2}
\begin{aligned}
&\mathcal{F}(E)=-\frac{E}{24\pi^2}\int^{\infty}_{-\infty}\,\mathrm{d}u\left[\chi^2(u)(E^2+\ddot{\x}^2)+3\dot{\chi}^2(u)\right]\\
&+\lim_{\eta\to 0}\frac{1}{2\pi^3}\int^{\infty}_{-\infty}\,\mathrm{d}u
\, 
\chi(u)\int^{\infty}_{\eta}\,\mathrm{d}s\Bigg(\frac{\chi(u-s)\cos{(Es)}}{\left[(\Delta \x)^2\right]^2}
-\frac{\chi(u)}{s^4}+\frac{\chi(u) (3E^2+\ddot{\x}^2)}{6s^2}-\frac{\ddot{\chi}(u)}{2s^2}\Bigg). 
\end{aligned}
\end{equation}

To take the limit $\eta\to0$ in~\eqref{eq:6d:result2}, 
we add and 
subtract under the $s$-integral terms that disentangle 
the small $s$ divergences of 
$\cos(Es)/\left[(\Delta \x)^2\right]^2$ from the small 
$s$ behaviour of $\chi(u-s)$. 
Following \cite{satz:smooth}, we obtain 
\begin{equation}
\begin{aligned}
\label{eq:6d:detresp}
\mathcal{F}(E) & =
-\frac{E}{24\pi^2}\int^{\infty}_{-\infty}\,\mathrm{d}u
\left[
\chi^2(u)
\left(E^2+\ddot{\x}^2\right)
+3\dot{\chi}^2(u)
\right]
\\
&\ -\frac{E^2}{4\pi^3}\int^{\infty}_{0}\,\frac{\mathrm{d}s}{s^2}\int^{\infty}_{-\infty}\,\mathrm{d}u\, \chi(u)\left[\chi(u-s)-\chi(u)\right]\\
&\ +\frac{1}{2\pi^3}\int^{\infty}_{0}\,\frac{\mathrm{d}s}{s^4}\int^{\infty}_{-\infty}\,\mathrm{d}u\, \chi(u)
\bigl[
\chi(u-s)-\chi(u)
-\tfrac12 s^2\ddot{\chi}(u)
\bigr]
\\
&\ -\frac{1}{12\pi^3}\int^{\infty}_{0}\,\frac{\mathrm{d}s}{s^2}
\int^{\infty}_{-\infty}\,\mathrm{d}u\,\chi(u)
\Bigl\{
\left[\chi(u-s)-\chi(u)\right] \ddot{\x}^2
-s \chi(u-s) \, \ddot{\x}\cdot\x^{(3)}
\Bigr\}
\\
&\ +\frac{1}{2\pi^3}\int^{\infty}_{-\infty}\,\mathrm{d}u\,\chi(u)\int^{\infty}_{0}\,\mathrm{d}s\,\chi(u-s)\left(\frac{\cos{(Es)}}{\left[(\Delta \x)^2\right]^2}
-\frac{1}{s^4}+\frac{3E^2+\ddot{\x}^2}{6s^2}-\frac{\ddot{\x}\cdot\x^{(3)}}{6s}\right)
\ , 
\end{aligned}
\end{equation}
where 
$\ddot{\x}^2$ and $\ddot{\x}\cdot\x^{(3)}$ are evaluated at~$u$. 
The interchanges of the integrals before taking 
the limit $\eta\to0$ are 
justified by absolute convergence of the double integrals, 
and taking the limit $\eta\to0$ under the outer integral is 
justified by dominated convergence: 
in each integral over $s$ in~\eqref{eq:6d:detresp},  
the integrand is regular as $s\to0$. 

Equation \eqref{eq:6d:detresp} is our final, 
regulator-free expression for the 
response function. In section \ref{sec:sharp}
we shall consider its behaviour when the switching approaches the step-function.

\section{Response function for $d=3$}
\label{sec:3d}

In this section we remove the regulator from the response function 
formula \eqref{eq:respfunc-alt} for $d=3$.
The qualitatively new feature is that the techniques of 
section \ref{sec:6d} 
need to be adapted to the fractional power in the Wightman function. 

\subsection{Regularisation}
\label{sec:3d:reg}

The regularised $d=3$ Wightman function 
reads \cite{kay-wald,Decanini:2005gt,Langlois-thesis}
\begin{equation}
\label{eq:w3d:1}
W_\epsilon (u,u-s)=\frac{1}{4\pi}\frac{1}{\sqrt{{(\Delta\x)}^2+2i\epsilon\Delta t+\epsilon^2}}
\ , 
\end{equation}
where the branch of the square root is such that 
the $\epsilon\to0$ limit of the square root is positive when ${(\Delta\x)}^2>0$. 
Separating the real and imaginary parts gives 
\begin{equation}
\label{eq:w3d:3}
W_\epsilon (u,u-s) = 
\frac{\sqrt{R+{(\Delta\x)}^2+\epsilon^2}-i\sqrt{R-{(\Delta\x)}^2-\epsilon^2}}{4\sqrt{2}\,\pi \, R}
\end{equation}
where $R$ is given by~\eqref{eq:R-def}, 
the quantities under the square roots are positive 
and the square roots are taken positive. 
From \eqref{eq:respfunc-alt} we now obtain 
\begin{equation}
\begin{aligned}
\label{eq:3d:detresp}
\mathcal{F}(E)
&=
\lim_{\epsilon\to 0}\frac{1}{2\sqrt{2}\,\pi} \int^{\infty}_{-\infty}\,\mathrm{d}u\,\chi (u)\int^{\infty}_{0}\,\mathrm{d}s\,
\frac{\chi(u-s)}{R}\times 
\\[1ex]
& \hspace{4ex}
\times \left[\sqrt{R+{(\Delta\x)}^2+\epsilon^2}\cos{(Es)}-\sqrt{R-{(\Delta\x)}^2-\epsilon^2}\sin{(Es)}\right]
\ . 
\end{aligned}
\end{equation}

We proceed as in section~\ref{sec:6d}. 
Working under the expression
${\left(2\sqrt{2}\,\pi\right)}^{-1}
\int^{\infty}_{-\infty}\,\mathrm{d}u\,\chi (u)$, we write the integral
over $s$ as the sum $I^{\text{even}}_{<} + I^{\text{odd}}_{<} +
I^{\text{even}}_{>} + I^{\text{odd}}_{>}$, where the notation follows
section \ref{sec:6d} with the exception that we now choose
$\eta:=\epsilon^{1/2}$. We consider the two intervals of $s$ in the
next two subsections. 

\subsection{Subinterval $\eta < s < \infty$}
\label{sec:3d:larges}

Consider $I^{\text{odd}}_{>}$. 
When $\epsilon$ is set to zero, the integrand in 
$I^{\text{odd}}_{>}$ reduces to 
$- \chi(u-s)\sin{(Es)}\sqrt{-2/{(\Delta\x)}^2}$, 
where the quantity under the square root 
is positive and the square root is taken positive. 
This replacement creates in $I^{\text{odd}}_{>}$
an error that can be arranged in the form  
\begin{equation} 
\begin{aligned}
\label{eq:3d:error:2}
&\int^{\infty}_{\eta}\,\mathrm{d}s\,\frac{\chi(u-s)\epsilon^2}{{(\Delta\x)}^2\sqrt{-{(\Delta\x)}^2}}\times
\\[1ex]
&\hspace{6ex}
\times \left[\frac{\left(2+\frac{\epsilon^2}{(\Delta\x)^2}+4\frac{(\Delta t)^2}{(\Delta\x)^2}-2\frac{S}{(\Delta\x)^2}-\epsilon^2\frac{S^2}{\left[(\Delta\x)^2\right]^3}\right)}{\frac{R}{(\Delta\x)^2}
\left(\sqrt{1+\frac{\epsilon^2}{(\Delta\x)^2}-\frac{R}{(\Delta\x)^2}}-\frac{\sqrt{2}R}{(\Delta\x)^2}\right)
\left(\frac{R}{(\Delta\x)^2}-1-\frac{\epsilon^2S}{\left[(\Delta\x)^2\right]^2}\right)} \right] 
\ , 
\end{aligned}
\end{equation} 
where $S := 3{(\Delta\x)}^2+2\epsilon^2+8{(\Delta t)}^2$. 

Using bounding arguments similar to those in section~\ref{sec:6d}, 
we find that $R/(\Delta\x)^2=- 1 + O\left(\eta^2\right)$, 
$\epsilon^2/(\Delta\x)^2=O\left(\eta^2\right)$
and 
$(\Delta t)^2/(\Delta\x)^2=O\left(1\right)$, 
and as a consequence 
$S/(\Delta\x)^2=O\left(1\right)$. 
The integrand in \eqref{eq:3d:error:2}
is hence bounded in absolute value by a constant times 
$\epsilon^{2}/\left[-{(\Delta\x)}^2\right]^{3/2} \le \epsilon^{2}/s^3$, from 
which it follows that the integral is of order 
$O\left(\epsilon^{2}/\eta^2\right) = O\left(\eta^2\right)$. 

Similar estimates show that $I^{\text{even}}_{>}=O\left(\eta\right)$. 

Collecting, we have 
\begin{equation}
\label{eq:3d:allupper}
I^{\text{even}}_{>}+I^{\text{odd}}_{>}=-\int^{\infty}_{\eta}\,\mathrm{d}s\,
\chi(u-s)\sin{(Es)}\sqrt{\frac{-2}{{(\Delta\x)}^2}}
\ \ +\ O\left(\eta\right)
\ . 
\end{equation}

\subsection{Subinterval $0 < s < \eta$}
\label{sec:3d:smalls}

Consider $I^{\text{odd}}_{<}$, given by 
\begin{equation}
\label{eq:3d:lessodd:1}
I^{\text{odd}}_{<}=-\int^{\eta}_{0}\,\mathrm{d}s\,\chi(u-s)
\, \frac{\sin{(Es)}}{R}\sqrt{R-{(\Delta\x)}^2-\epsilon^2}
\ . 
\end{equation}
Writing $s = \epsilon r$
and introducing the book-keeping parameter $z$ as in 
section~\ref{sec:6d}, 
the counterpart of \eqref{eq:den6d} reads 
\begin{equation}
\begin{aligned}
\label{eq:den3d}
R^2 
&= 
\epsilon^{4}P\Bigg[1-\frac{4\dot{t}\ddot{t}\epsilon r^3}{P}z
+\sum^{n_4-1}_{n=0}\left(2X_{(n+4)}+4T_{(n+4)}\right)\frac{\epsilon^{n+2}r^{n+4}}{P}z^{2+(n/2)}
\\[1ex]
& 
+\sum^{n_5-1}_{n=0}F_{(n+6)}\frac{\epsilon^{n+2}r^{n+6}}{P}z^{1+(n/2)}
+ \frac{O\bigl(\epsilon^{n_4+2}r^{n_4+4}\bigr)}{P}z^{2+(n_4/2)}
+ \frac{O\bigl(\epsilon^{n_5+2}r^{n_5+6}\bigr)}{P}z^{1+(n_5/2)}
\Bigg] \ , 
\end{aligned}
\end{equation}
where the powers of $z$ differ from those in 
\eqref{eq:den6d} because the range of $r$ is now 
$0\le r \le \epsilon^{-1/2}$. 
It follows that in the denominator in 
\eqref{eq:3d:lessodd:1}
we have $R = \epsilon^2 \sqrt{P} \left[1 + O(z)\right]$, and in the numerator we have $\chi(u-s) \sin(Es) = 
\epsilon E r \left[\chi(u) +  O\left(\sqrt{z}\right) \right]$. 

To estimate the square root in the 
numerator in~\eqref{eq:3d:lessodd:1},
we note first that all the terms with a positive power of $z$ in 
\eqref{eq:den3d} are at small $r$ asymptotic to a power of $r$ that is greater than~2. It follows that the same powers of $z$ are retained if these terms are multiplied by any positive function of $r$ that is bounded at small $r$ by a constant times $r^{-3}$ and at large $r$ by a constant. 

Now, we rearrange the quantity under the square root 
in \eqref{eq:3d:lessodd:1} as 
\begin{align}
R-{(\Delta\x)}^2-\epsilon^2 
& = 
\epsilon^2 Q \Biggl[1 
-\sum^{n_6-1}_{n=0} X_{(n+4)} \frac{\epsilon^{n+2} r^{n+4}}{Q}
+ \frac{\sqrt{P}}{Q}
\left(\frac{R}{\epsilon^2 \sqrt{P}} \, -1\right)
\nonumber
\\[1ex]
& \hspace{8ex}
+ \frac{O\bigl(\epsilon^{n_6+2} r^{n_6+4}\bigr)}{Q}
\Biggr]
\ , 
\label{eq:3d:tricky}
\end{align}
where 
\begin{align}
Q :=  \sqrt{P} + r^2 - 1
\end{align}
and the positive integer $n_6$ may be chosen arbitrarily. 
Note that $Q$ is positive for $r>0$, 
its small $r$ behaviour is $Q = 2{\dot{t}}^2 r^2 + O\left(r^4\right)$ 
where the coefficient of $r^2$ is positive, and its behaviour at large $r$ 
is 
$Q/r^2 = 2+ O\left(r^{-2}\right)$. 
We wish to regard the external factor $\epsilon^2 Q$ 
in \eqref{eq:3d:tricky} as the dominant part and the 
terms in the square brackets as a leading 1 plus 
sub-leading corrections. In the terms proportional to  
$X_{(n+4)}$, this is accomplished by 
inserting the book-keeping factors  
$z^{1 + (n/2)}$. From the asymptotic behaviour of $\sqrt{P}/Q$ 
at small and large $r$ we see that in the 
term involving $\sqrt{P}/Q$ 
this is accomplished by taking $R^2$ to be given by~\eqref{eq:den3d}, with the $z$-factors therein. 
A~Taylor expansion in $z^{1/2}$ then shows that $R-{(\Delta\x)}^2-\epsilon^2 = \epsilon^2 Q\left[1 + O(z)\right]$. 

Collecting, we find 
\begin{align}
I^{\text{odd}}_{<}
& =
-\epsilon E\int^{1/\eta}_0\,\mathrm{d}r\,\frac{r \sqrt{Q}}{\sqrt{P}} 
\left[\chi(u)+O\left(\sqrt{z}\,\right)\right] 
\notag
\\
\label{eq:3d:loweroddfinal}
& =O\left(\eta\right)
\ , 
\end{align}
where the final form follows because the integrand 
asymptotes to a constant at large~$r$. 

Consider then $I^{\text{even}}_{<}$, given by 
\begin{equation}
\label{eq:3d:lesseven}
I^{\text{even}}_{<}=
\int^{\eta}_{0}\,\mathrm{d}s\,\chi(u-s)
\, \frac{\cos{(Es)}}{R}\sqrt{R+{(\Delta\x)}^2+\epsilon^2}
\ . 
\end{equation}
We now rearrange the quantity under the square root in 
\eqref{eq:3d:lesseven} as 
\begin{align}
R+{(\Delta\x)}^2+\epsilon^2 
& = 
\epsilon^2 N \Biggl[1 
+\sum^{n_7-1}_{n=0} X_{(n+4)} \frac{\epsilon^{n+2} r^{n+4}}{N}
+ \frac{\sqrt{P}}{N}
\left(\frac{R}{\epsilon^2 \sqrt{P}} \, -1\right)
\nonumber
\\[1ex]
& 
\hspace{8ex} 
+ \frac{O\bigl(\epsilon^{n_7+2} r^{n_7+4}\bigr)}{N}
\Biggr]
\ , 
\label{eq:3d:tricky2}
\end{align}
where 
\begin{align}
N :=  \sqrt{P} + 1 - r^2
\end{align}
and the positive 
integer $n_7$ may be chosen arbitrarily. 
Note that $N$ is positive, 
its small $r$ behaviour is $N = 2 + O\left(r^2\right)$ 
and its large $r$ 
behaviour is 
$N = 2{\dot{t}}^2 + O\left(r^{-2}\right)$. 
We wish to regard the external factor $\epsilon^2N$ 
in \eqref{eq:3d:tricky2} as the dominant part. 
In the square brackets, the terms proportional 
to $X_{(n+4)}$ can be given the
book-keeping factors  
$z^{n/2}$, 
while in the term involving $\sqrt{P}/N$, the large $r$ behaviour of 
$\sqrt{P}/N$ implies that the powers of $z$ inherited from 
\eqref{eq:den3d} must be appropriately decreased. 
Using $F_6 = - 2 X_4$, we find 
\begin{align}
\label{eq:3d:tricky3}
R+{(\Delta\x)}^2+\epsilon^2 
= 
\epsilon^2 N \left[1 
+X_4 \frac{\epsilon^{2} r^{4}}{N}
\left(1 - \frac{r^2}{\sqrt{P}}\right)
+ O\left(\sqrt{z}\right)
\right]
\ . 
\end{align}
Although the term proportional to $X_4$ in the square brackets in 
\eqref{eq:3d:tricky3} has arisen as a combination of two individual terms that came with $z$-factors $z^0$, 
a cancellation between these individual terms at large $r$ implies that the term as a whole can now be reassigned the factor~$z$. We hence have 
$\sqrt{R+{(\Delta\x)}^2+\epsilon^2} 
= 
\epsilon \sqrt{N} \left[1 
+ O\left(\sqrt{z}\right)
\right]$. 
Using this and \eqref{eq:den3d} in $\eqref{eq:3d:lesseven}$, 
we obtain 
\begin{align}
I^{\text{even}}_{<}
& = \int^{1/\eta}_0\,\mathrm{d}r\, \frac{\sqrt{N}}{\sqrt{P}}
\left[\chi(u) +O\left(\sqrt{z}\,\right) \right]
\notag
\\[1ex]
\label{eq:3d:lowerevenfinal}
&=
\frac{\pi\chi(u)}{\sqrt{2}} +O\left(\eta\right)
\ , 
\end{align}
where the final 
form comes by 
evaluating the elementary integral. 

\subsection{Combining the subintervals}

Combining (\ref{eq:3d:allupper}), (\ref{eq:3d:loweroddfinal}) and 
(\ref{eq:3d:lowerevenfinal}), 
we obtain from \eqref{eq:3d:detresp}
for the response function the final form 
\begin{align}
\label{eq:3d:detrespfinal}
\mathcal{F}\left(E\right)
=
\frac{1}{4}\int^{\infty}_{-\infty}\,\mathrm{d}u\,\chi^2(u)
\ - \ 
\frac{1}{2\pi}\int^{\infty}_{-\infty}\,\mathrm{d}u\,\chi(u)\int^{\infty}_{0}\,\mathrm{d}s \, 
\frac{\chi(u-s) \sin(Es)}{\sqrt{-{(\Delta\x)}^2}}
\ .  
\end{align}
The limit $\eta\to0$ has been taken by just setting the lower limit of the $s$-integral to zero, as the small $s$ behaviour of the numerator cancels the singularity in the denominator.

\section{Response function for $d=5$}
\label{sec:5d}

In this section we remove the regulator from the response function 
formula \eqref{eq:respfunc-alt} for $d=5$, extending 
the technique of section~\ref{sec:3d}. 

\subsection{Regularisation}
\label{sec:5d:reg}

The regularised $d=5$ Wightman function 
reads \cite{kay-wald,Decanini:2005gt,Langlois-thesis}
\begin{equation}
\label{eq:w5d:1}
W_\epsilon (u,u-s)=\frac{1}{8\pi^2}\frac{1}{{\left[{(\Delta\x)}^2+2i\epsilon\Delta t+\epsilon^2\right]}^{3/2}}
\ . 
\end{equation}
Separating the real and imaginary parts and substituting in 
\eqref{eq:respfunc-alt} yields 
\begin{equation}
\begin{aligned}
\label{eq:5d:detres}
&\mathcal{F}\left(E\right)=\lim_{\epsilon\to 0}
\frac{1}{4\sqrt{2}\,\pi^2}\int^{\infty}_{-\infty}\,\mathrm{d}u\,\chi(u)\,\int^{\infty}_{0}\,\mathrm{d}s\,\frac{\chi(u-s)}{R^3} \, \times\\
&\times 
\Bigg[\cos(Es) 
\left( \bigl({(\Delta\x)}^2+\epsilon^2\bigr)\sqrt{R+{(\Delta\x)}^2+\epsilon^2}-2\epsilon(\Delta t)\sqrt{R-{(\Delta\x)}^2-\epsilon^2} \, \right) 
\\
&\hspace{4ex}
-\sin(Es) 
\left( \bigl({(\Delta\x)}^2+\epsilon^2\bigr)\sqrt{R-{(\Delta\x)}^2-\epsilon^2}+2\epsilon(\Delta t)\sqrt{R+{(\Delta\x)}^2+\epsilon^2} \, \right)\Bigg]
\ . 
\end{aligned}
\end{equation}
Working under the expression 
${(4\sqrt{2}\,\pi^2)}^{-1} \int^{\infty}_{-\infty}\,\mathrm{d}u\,\chi(u)$, we again write the integral 
over $s$ as the sum $I^{\text{even}}_{<} + I^{\text{odd}}_{<} +
I^{\text{even}}_{>} + I^{\text{odd}}_{>}$, 
choosing $\eta:=\epsilon^{1/4}$ as in section~\ref{sec:6d}. 

\subsection{Subinterval $\eta < s < \infty$}
\label{sec:5d:larges}

$I^{\text{odd}}_{>}$ and $I^{\text{even}}_{>}$
can be estimated as in section~\ref{sec:3d}. 
We merely record here the outcome, 
\begin{align}
\label{eq:5d:upper}
I^{\text{odd}}_{>}+I^{\text{even}}_{>}=\int^{\infty}_{\eta}\,\mathrm{d}s\,\chi(u-s)\sin{(Es)}\sqrt{\frac{-2}{\left[{(\Delta\x)}^2\right]^3}}
\ \ +\ O\left(\eta\right)
\ , 
\end{align}
where the integral term has arisen from replacing 
the integrand in $I^{\text{odd}}_{>}$ 
by its pointwise $\epsilon\to0$ limit. 

\subsection{Subinterval $0 < s < \eta$}
\label{sec:5d:smalls}

Consider $I^{\text{even}}_{<}$ 
and $I^{\text{odd}}_{<}$, given by 
\begin{subequations}
\label{eq:5dlessboth}
\begin{align}
I^{\text{even}}_{<}&=\int^{\eta}_{0}\,\mathrm{d}s\,\chi(u-s)\frac{\cos{(Es)}}{R^3} \, \times 
\notag 
\\[1ex]
& \hspace{2ex}
\times 
\left[ 
\bigl({(\Delta\x)}^2+\epsilon^2\bigr)\sqrt{R+{(\Delta\x)}^2+\epsilon^2}
-2\epsilon
(\Delta t)\sqrt{R-{(\Delta\x)}^2-\epsilon^2} \, 
\right]
\ , 
\label{eq:5dlesseven}
\\
I^{\text{odd}}_{<}&= 
- \int^{\eta}_{0}\,\mathrm{d}s\,\chi(u-s)\frac{\sin{(Es)}}{R^3} \, \times 
\notag 
\\[1ex]
& \hspace{4ex}
\times 
\left[ 
\bigl({(\Delta\x)}^2+\epsilon^2\bigr)\sqrt{R-{(\Delta\x)}^2-\epsilon^2}
+2\epsilon
(\Delta t)\sqrt{R+{(\Delta\x)}^2+\epsilon^2} \, 
\right]
\ .
\label{eq:5dlessodd}
\end{align}
\end{subequations}
In the $R^3$ in the denominator, we use~\eqref{eq:den6d}. 
In the square root 
$\sqrt{R-{(\Delta\x)}^2-\epsilon^2}$ in the numerator, 
we use~\eqref{eq:3d:tricky}, 
inserting the factors $z^{(n+2)/4}$ 
in the terms proportional to $X_{(n+4)}$ and using in the last term 
\eqref{eq:den6d} for~$R$: 
by the asymptotic behaviour of $\sqrt{P}/Q$ and the observations made in section~\ref{sec:3d}, this makes $z$ into an appropriate parameter for organising the square brackets in \eqref{eq:3d:tricky} 
into a Taylor expansion in $z^{1/4}$ with the leading term~$1$. 

In the square root 
$\sqrt{R+{(\Delta\x)}^2+\epsilon^2}$ in the numerator, 
we wish to use~\eqref{eq:3d:tricky2}. 
Attempting to regard the $1$ in the square brackets as the dominant term can at first sight seem problematic because the terms 
proportional to $X_{(n+4)}$ acquire the $z$-factors 
$z^{-1 + (n/4)}$, where the exponent is nonpositive for $n\le 4$, and 
when the last term is Taylor expanded in $z^{1/4}$ 
using~\eqref{eq:den6d}, the asymptotic behaviour of the factor 
$\sqrt{P}/N$ implies that the exponents of $z$ must be appropriately decreased and some of these decreased exponents are nonpositive. 
However, the nonpositive powers of $z$ coming from the last term and from the terms proportional to $X_{(n+4)}$ 
can be grouped into combinations that can be reassigned 
positive powers of~$z$, similarly to what happened for $d=3$ in~\eqref{eq:3d:tricky3}. 
After these reassignments we therefore obtain for 
$\sqrt{R+{(\Delta\x)}^2+\epsilon^2}$ a Taylor expansion in $z^{1/4}$
that starts as 
$\epsilon\sqrt{N}\left[1+O\left(z^{1/4}\right)\right]$. 

The remaining factors in \eqref{eq:5dlessboth} 
are expanded in $z$ as in section~\ref{sec:6d}. 

We can now perform the Taylor expansion in~$z^{1/4}$ under the integrals. 
Dropping powers of $z$ that are too high to contribute in the $\epsilon\to0$ limit, and setting then $z=1$, we 
obtain for 
$I^{\text{even}}_{<}$ 
and $I^{\text{odd}}_{<}$ formulas that consist of 
sums of finitely many elementary integals 
plus an error term that vanishes in the $\epsilon\to0$ limit. 
The elementary integrals are 
of the form 
$\int^{\eta^{-3}}_{0}\,\mathrm{d}r\,r^n N^{\pm 1/2} P^{-m}$, 
$\int^{\eta^{-3}}_{0}\,\mathrm{d}r\,r^n N^{-3/2} P^{-m}$, 
$\int^{\eta^{-3}}_{0}\,\mathrm{d}r\,r^n Q^{\pm 1/2} P^{-m}$ 
and 
$\int^{\eta^{-3}}_{0}\,\mathrm{d}r\,r^n Q^{-3/2} P^{-m}$, 
where $n$ is a positive integer and $m$ is a positive 
integer or half-integer.

\subsection{Combining the subintervals}

Evaluating the numerous elementary integrals 
that came from \eqref{eq:5dlessboth}, 
combining the results with 
\eqref{eq:5d:upper} 
and proceeding as in section~\ref{sec:6d}, we 
find from 
\eqref{eq:5d:detres}
that the response function is given by  
\begin{align}
\mathcal{F}\left(E\right)
&=\frac{1}{64\pi}\int^{\infty}_{-\infty}\,\mathrm{d}u\,\left[\chi^2\left(4E^2+\ddot{\x}^2\right)+4\dot{\chi}^2\right]
\notag
\\
&\hspace{2ex}
+\lim_{\eta\to 0}\frac{1}{4\pi^2}\int^{\infty}_{-\infty}\,\mathrm{d}u \, \chi(u)\,\int^{\infty}_{\eta}\,\mathrm{d}s\,
\left(
\frac{\chi(u-s) \sin{\left(Es\right)}}{\sqrt{\bigl[-{(\Delta\x)}^2\bigr]^3}}-\frac{E\chi(u)}{s^2}
\right)
\ , 
\end{align}
To take the limit 
$\eta\to0$, 
we add and subtract under the $s$-integral terms that 
disentangle the small $s$ divergence of 
$\sin{\left(Es\right)} \bigl[-{(\Delta\x)}^2\bigr]^{-3/2}$ 
from the small $s$ behaviour of $\chi(u-s)$. 
Proceeding as in section~\ref{sec:6d}, we find 
\begin{equation}
\begin{aligned}
\label{eq:5d:detresp:ssform}
\mathcal{F}\left(E\right)&=\frac{1}{64\pi}\int^{\infty}_{-\infty}\,\mathrm{d}u\,\left[\chi^2(u)\left(4E^2+\ddot{\x}^2\right)+4\dot{\chi}^2(u)\right]
\\[1ex]
&\ +\frac{E}{4\pi^2}\int^{\infty}_{0}\,\frac{\mathrm{d}s}{s^2}\int^{\infty}_{-\infty}\,\mathrm{d}u\, \chi(u)\left[\chi(u-s)-\chi(u)\right]\\
&\ +\frac{1}{4\pi^2}\int^{\infty}_{-\infty}\,\mathrm{d}u\, \,\chi(u) 
\int^{\infty}_{0}\,\mathrm{d}s\,\chi(u-s)
\left(\frac{\sin{(Es)}}{\sqrt{\left[-{(\Delta\x)}^2\right]^3}}-\frac{E}{s^2}\right)
\ . 
\end{aligned}
\end{equation}

\section{Sharp switching limit}
\label{sec:sharp}

In this section we consider the limit in which the switching function approaches a step-function of unit height and fixed duration. Concretely, we take \cite{satz:smooth,satz-louko:curved}
\begin{equation}
\label{eq:ss:switchingfunc}
\chi(u)=h_1\left(\frac{u-\tau_0+\delta}{\delta}\right)
\times 
h_2\left(\frac{-u+\tau+\delta}{\delta}\right)
\ , 
\end{equation}
where the parameters 
$\tau$, $\tau_0$ and $\delta$ satisfy 
$\tau>\tau_0$ and $\delta>0$, 
and 
$h_1$ and $h_2$ are smooth non-negative functions satisfying 
$h_i(x)=0$ for $x\le0$ and $h_i(x)=1$ for $x\ge1$. 
In words, the detector
is switched on over an 
interval of duration $\delta$ just before 
proper time~$\tau_0$, 
it stays on until proper time~$\tau$, 
and it is switched off over an interval of duration 
$\delta$ just after proper time~$\tau$. 
The manner of the switch-on and switch-off 
is specified respectively by the functions $h_1$ and~$h_2$. 
The limit of sharp switching is then $\delta\to0$, 
with $\tau_0$ and $\tau$ fixed. 

We denote the response function by~$\mathcal{F}_\tau$, 
where the subscript serves as an explicit reminder of the dependence on the switch-off moment~$\tau$. We are interested both in $\mathcal{F}_\tau$ and in its derivative with respect to~$\tau$, which we denote by $\mathcal{\dot{F}}_{\tau}$. As mentioned in section~\ref{sec:intro}, 
$\mathcal{\dot{F}}_{\tau}$ can be regarded as the detector's instantaneous transition rate per unit proper time, observationally meaningful in terms of consequent measurements in identical ensembles of detectors~\cite{satz-louko:curved}. 

The case of two-dimensional Minkowski spacetime, $d=2$, 
was discussed in section~\ref{sec:detectormodels}. 
We shall address the cases from 
$d=3$ to $d=6$ in the following subsections. 

\subsection{$d=3$}

For $d=3$, $\mathcal{F}_\tau$ is given by~\eqref{eq:3d:detrespfinal}. 
The limit $\delta\to0$ is well defined and 
can be taken directly in~\eqref{eq:3d:detrespfinal}, 
with the result 
\begin{align}
\label{eq:ss:3d:detrespSS}
\mathcal{F}_\tau (E)
= 
\frac{\Delta\tau}{4}
-\frac{1}{2\pi}\int^{\tau}_{\tau_0}\,\mathrm{d}u\,\int^{u-\tau_0}_{0}\,\mathrm{d}s\,\frac{\sin{(Es)}}{\sqrt{-{(\Delta\x)}^2}}
\ , 
\end{align}
and differentiation with respect to $\tau$ gives 
\begin{align}
\label{eq:ss:3d:transrate}
\dot{\mathcal{F}}_{\tau}\left(E\right)
=
\frac{1}{4}-\frac{1}{2\pi}\int^{\Delta\tau}_{0}\,\mathrm{d}s\,\frac{ \sin{\left(Es\right)}}{\sqrt{-\left(\Delta\x\right)^2}}
\ . 
\end{align}

\subsection{$d=4$}

The case $d=4$ was addressed in~\cite{satz:smooth}. 
The expression for the response function with 
a general switching function reads 
\begin{align}
\mathcal{F}(E)
&= -\frac{E}{4\pi}\int_{-\infty}^{\infty}\mathrm{d}u\,\chi^2(u)\,
+ \frac{1}{2\pi^2}\int_0^{\infty}\frac{\mathrm{d}s}{s^2}\int_{-\infty}^{\infty}\mathrm{d}u\,\chi(u)\left[ \chi(u)-\chi(u-s)\right]
\notag
\\[1ex]
& 
\hspace{3ex}
+\frac{1}{2\pi^2}\int_{-\infty}^{\infty}\mathrm{d}u\,\chi(u)\int_0^{\infty}\mathrm{d}s\,\chi(u-s)\left( \frac{\cos(Es)}{{(\Delta \x)}^2}+\frac{1}{s^2}\right) 
\ . 
\label{eq:4dFraw}
\end{align}
The first and third terms in \eqref{eq:4dFraw} have well-defined limits as $\delta\to0$.  
The second term in \eqref{eq:4dFraw} takes at small $\delta$ the form ${(2\pi^2)}^{-1} \ln(\Delta\tau/\delta) + C + O(\delta/\Delta\tau)$, where $C$ is a constant determined by the functions $h_1$ and~$h_2$, and this term hence diverges logarithmically as $\delta\to0$. However, the $\tau$-derivative of this term remains finite as $\delta\to0$, and 
the transition rate has the well-defined limit 
\begin{align}
\label{eq:ss:4d:transrate}
\dot{\mathcal{F}}_{\tau}\left(E\right)
=
-\frac{E}{4\pi}+\frac{1}{2\pi^2}
\int_0^{\Delta\tau}\textrm{d}s
\left( 
\frac{\cos (Es)}{{(\Delta \x)}^2} 
+ 
\frac{1}{s^2} 
\right) 
\ \ +\frac{1}{2\pi^2 \Delta \tau}
\ . 
\end{align}

\subsection{$d=5$}

For $d=5$, 
$\mathcal{F}_\tau$ is given by~\eqref{eq:5d:detresp:ssform}. 
The last term in \eqref{eq:5d:detresp:ssform} has a well-defined 
limit as $\delta\to0$, and so has the part of the 
first term containing~$\chi^2$. 
The part of the first term containing 
${\dot\chi}^2$ equals~$C'/\delta$, 
where $C'$ is a positive constant 
determined by the functions $h_1$ and~$h_2$. 
This part diverges as $\delta\to0$ but is 
independent of $\tau$ and does therefore 
not contribute to~$\dot{\mathcal{F}}_{\tau}$. 
Finally, the second term in \eqref{eq:5d:detresp:ssform}
is similar to the second term in the $d=4$ formula 
\eqref{eq:4dFraw}, being logarithmically divergent as $\delta\to0$ 
but having a $\tau$-derivative that has a well-defined limit as 
$\delta\to0$. 

Collecting, we find that the transition rate has a 
well-defined $\delta\to0$ limit, given by  
\begin{align}
\label{eq:ss:5d:transrate}
\dot{\mathcal{F}}_{\tau}\left(E\right)
=
\frac{4E^2 + \ddot{\x}^2(\tau)}{64\pi}
+ 
\frac{1}{4\pi^2}\int^{\Delta\tau}_{0}\,\mathrm{d}s
\left(\frac{\sin{(Es)}}{\sqrt{\left[-{(\Delta\x)}^2\right]^3}}-\frac{E}{s^2}\right)
\ -\frac{E}{4\pi^2\Delta\tau}
\ . 
\end{align}

\subsection{$d=6$}

For $d=6$, 
$\mathcal{F}_\tau$ is given by~\eqref{eq:6d:detresp}.
The last term in \eqref{eq:6d:detresp} remains finite as 
$\delta\to0$. The first and second terms are similar to those encountered in $d=5$, with contributions that diverge in the $\delta\to0$ limit 
proportionally to $1/\delta$ and $\ln\delta$, but with $\tau$-derivatives that remain finite in this limit. 

The third and fourth terms can be handled by breaking the integrations into subintervals as in~\cite{satz:smooth}. 
The third term diverges proportionally to $\delta^{-2}$ as $\delta\to0$, but its $\tau$-derivative has a well-defined limit as 
$\delta\to0$. 
The fourth term resembles the second term in that the divergence at $\delta\to0$ is logarithmic in~$\delta$, but the presence of 
$\ddot{\x}^2$ and $\ddot{\x}\cdot\x^{(3)}$ in the integrand has the consequence that the coefficient of the divergent logarithm depends on the trajectory and does not vanish on differentiation with respect to~$\tau$. 
We find that the transition rate is given by 
\begin{align}
\mathcal{\dot{F}}_{\tau}(E)
&=
\frac{\ddot{\x}(\tau)\cdot\x^{(3)}(\tau)}{12\pi^3}\left(\ln{\left(\frac{\Delta\tau}{\delta}\right)+C^{'}_{+}}\right)
-\frac{E\bigl(E^2+\ddot{\x}^2(\tau)\bigr)}{24\pi^2}
\notag
\\[1ex]
&\hspace{2ex}
+\frac{1}{2\pi^3}\int^{\Delta\tau}_{0}\,\mathrm{d}s\,\left(
\frac{\cos{(Es)}}{\left[(\Delta \x)^2\right]^2}
-\frac{1}{s^4}+\frac{3E^2+\ddot{\x}^2(\tau)}{6s^2}
-\frac{\ddot{\x}(\tau)\cdot\x^{(3)}(\tau)}{6s}\right)
\notag
\\[1ex]
&\hspace{2ex}
+\frac{3E^2+\ddot{\x}^2(\tau)}{12\pi^3\Delta\tau}
-\frac{1}{6\pi^3\Delta\tau^3}
+O\left(\delta\ln{\left(\frac{\Delta\tau}{\delta}\right)}\right)
\ , 
\label{eq:ss:6d:transrate}
\end{align}
where the constant $C^{'}_{+}$ is determined 
by the switch-off function $h_2$ by 
\begin{align}
C^{'}_{+}&=-2\int^1_0\,\mathrm{d}r\,\frac{1}{r^2}
\left(\int^1_0\,\mathrm{d}v\,h_2(1-v)\left[h_2(1-v+r)-h_2(1-v)\right]-\tfrac{1}{2}r\right)
\notag 
\\
&\hspace{2ex}-2\int^1_0\,\mathrm{d}v\,h_2(v)\left[1-h_2(v)\right]
\ . 
\label{eq:Cprimeplus}
\end{align}

The qualitatively new feature is that the transition rate 
\eqref{eq:ss:6d:transrate} does not have a well-defined limit 
for generic trajectories as $\delta\to0$, 
because the coefficient of 
$\ddot{\x}(\tau)\cdot\x^{(3)}(\tau)$ diverges in this limit; 
further, even if $\delta$ is kept finite, 
the coefficient of this term depends on the details of the switch-off profile through the constant $C^{'}_{+}$~\eqref{eq:Cprimeplus}. 
The limit exists only for trajectories whose 
scalar proper acceleration $\sqrt{{{\ddot\x}}^2}$ is constant over the trajectory, 
in which case the coefficient of the divergent term 
in \eqref{eq:ss:6d:transrate} vanishes. 
Note that this special class includes all trajectories that are 
uniformly accelerated, in the sense of 
following an orbit of a timelike Killing vector.

\section{Spacetime dimension versus sharp switching}
\label{sec:consistencychecks}

We have found that the sharp switching limit of the detector 
response function becomes increasingly singular as the 
spacetime dimension $d$ increases from $2$ to~$6$. 
In this section we discuss further aspects of this singularity. 

First, we have seen that that the sharp switching limit of the response function diverges for~$d\ge4$. For $d=4$ and $d=5$ the divergent term is independent of the total detection time, and the limit of the instantaneous transition rate is still finite. For $d=6$, however, the instantaneous transition rate diverges for generic trajectories. We summarise this behaviour in Table~\ref{table:dimcomp}. 

\begin{table}[t]
\begin{center}
  \begin{tabular}{ | l | l | l  |}
    \hline
    $d$ & $\mathcal{F}_\tau$ & $\mathcal{\dot{F}}_{\tau} \vphantom{{\text{\Large A}}^A}$
    \\ \hline
    2 & finite & finite 
    \\ \hline
    3 & finite & finite 
    \\ \hline
    4 & $\ln\delta$ & finite 
    \\ \hline
    5 & $1/\delta$ & finite 
    \\ \hline
    6 & $1/\delta^2\vphantom{\text{\large$A^A$}}$ & $\ddot{\x}\cdot\x^{(3)} \ln\delta$ 
    \\ \hline
    \end{tabular}
\end{center}
\caption{The divergent pieces of the 
total transition probability $\mathcal{F}_\tau$ 
and the instantaneous transition rate $\mathcal{\dot{F}}_{\tau}$ 
for spacetime dimensions $d=2,\ldots,6$ 
in the sharp switching limit.\label{table:dimcomp}}
\end{table}

Second, we re-emphasise that when the Wightman distribution $W$ in 
\eqref{eq:respfunc-def} or \eqref{eq:respfunc-alt}
is represented as the $\epsilon\to0$ limit of the regularised Wightman function~$W_\epsilon$, 
the $\epsilon\to0$ limit needs to be taken \emph{before\/} considering the sharp switching limit: 
this is the only way one is guaranteed to be implementing the technical definition of the Wightman function correctly. 
With the regulator that we have used in this paper 
[equations
\eqref{eq:w6d}, 
\eqref{eq:w3d:1}
and~\eqref{eq:w5d:1}], it is known that 
attempting to reverse the limits na\"\i{}vely for $d=4$ 
would yield an incorrect, and even Lorentz-noncovariant, result for the transition rate for all noninertial trajectories~\cite{schlicht,Schlicht:thesis,louko-satz:profile}. 
We have verified that attempting to reverse the limits na\"\i{}vely 
would be incorrect also for $d=3$, $d=5$ and $d=6$. 
For $d=3$, substituting the 
regularised Wightman function 
\eqref{eq:w3d:1} 
in 
\eqref{eq:tranrate} 
and evaluating the limit by the method of section \ref{sec:3d} does 
give the correct result~\eqref{eq:ss:3d:transrate}, but attempting to take 
the limit $\epsilon\to0$ in \eqref{eq:tranrate} 
na\"\i{}vely under the integral would miss 
the first of the two terms in~\eqref{eq:ss:3d:transrate}. 
For $d=5$, substituting the regularised Wightman function 
\eqref{eq:w5d:1} in the na\"\i{}ve transition rate formula 
\eqref{eq:tranrate}
and evaluating the limit $\epsilon\to0$ 
by the methods of section \ref{sec:5d} yields for the transition 
rate an expression that consists of 
\eqref{eq:ss:5d:transrate} plus the Lorentz-noncovariant terms 
\begin{align}
\frac{\ddot{t}(2+\dot{t})E}{8\pi^2{(1+\dot{t})}^2}
-\frac{\ddot{t}}{8\pi^2{(1+\dot{t})}^2 \epsilon}
\ , 
\end{align}
of which the second diverges as $\epsilon\to0$. 
For $d=6$, starting with the regularised Wightman function 
\eqref{eq:w6d}
yields for the transition rate a formula that is similar
to~\eqref{eq:ss:6d:transrate}, 
with the logarithmically divergent term replaced by a 
term that is logarithmically divergent in~$\epsilon$, 
plus a number of Lorentz-noncovariant terms. 

Third, since the sharp switching divergence of $\mathcal{\dot{F}}_{\tau}$ for $d=6$ is perhaps surprising, we have verified that a similar divergence occurs also in the pointlike detector model where the switching is sharp at the outset but the detector is initially spatially smeared, having the Lorentz-function spatial profile with an overall size parameter~$\epsilon$, 
and the pointlike detector is recovered in the limit $\epsilon\to0$~\cite{schlicht,Langlois}. 
(The model can be alternatively regarded as that of a sharply-switched pointlike detector whose Wightman function is regularised in terms of the frequency measured in the detector's instantaneous rest frame, rather than in terms of the frequency measured in an externally-specified Lorentz frame~\cite{Langlois}.) Adapting the methods of section \ref{sec:6d} and proceeding as in~\cite{louko-satz:profile}, we find that the expression for $\mathcal{\dot{F}}_{\tau}$ is obtained from 
\eqref{eq:ss:6d:transrate} by the replacement 
$\ln(\tau/\delta) + C^{'}_{+} \to \ln(\tau/\epsilon) -\frac43 - \ln2$, 
so that the pointlike detector limit 
$\epsilon\to0$ is again divergent unless the trajectory has constant scalar acceleration. 

Fourth, for a trajectory of uniform linear acceleration~$a$, 
switched on in the inifinite past, 
the transition rate formulas 
\eqref{eq:ss:3d:transrate}, 
\eqref{eq:ss:4d:transrate}, 
\eqref{eq:ss:5d:transrate}
and 
\eqref{eq:ss:6d:transrate} 
yield 
\begin{align}
\label{eq:takagi:alld}
\begin{aligned}
&\dot{\mathcal{F}}_{d=3}\left(E\right)
=
\frac{1}{2}\frac{1}{e^{2\pi E/a}+1}
\ , \ \ \ 
&&
\dot{\mathcal{F}}_{d=5}\left(E\right)
=
\frac{1}{32\pi}\frac{\left(4 E^2+a^2 \right)}{e^{2\pi E/a}+1}\ , 
\\[1ex]
&\dot{\mathcal{F}}_{d=4}\left(E\right)
=
\frac{1}{2\pi}\frac{E}{e^{2\pi E/a}-1}
\ , \ \ \ 
&&\dot{\mathcal{F}}_{d=6}\left(E\right)
=\frac{1}{12\pi^2}\frac{E\left(E^2+a^2\right)}{e^{2\pi E/a}-1} 
\ . 
\end{aligned}
\end{align}
This was verified for $d=4$ in~\cite{louko-satz:profile}, 
and we have used the same contour deformation 
method for the other values of~$d$. 
The results \eqref{eq:takagi:alld} agree 
with those found in~\cite{takagi}, equation~(4.1.27), 
where they were obtained from 
a definition of transition rate that relies at the outset 
on the stationarity of the trajectory. 

Finally, we would like to speculate on how the response function and transition rate 
patterns that we have found for $d\le6$ might continue to $d>6$, and specifically to $d=7$. 

Recall that the formula \eqref{eq:respfunc-alt} gives the response function in terms of the distributional Wightman function~$W$.
If $W$ is to be replaced by the unregularised Wightman function under the integrals, then the negative powers of $s$ in 
$\Realpart \left[e^{-iEs} \, W(u,u-s) \right]$ 
must be subtracted. 
The last term in our formulas 
\eqref{eq:6d:detresp}, 
\eqref{eq:3d:detrespfinal}, 
\eqref{eq:5d:detresp:ssform}, 
and 
\eqref{eq:4dFraw}
is precisely of this form. 
The corresponding term can be constructed for any~$d$, 
and for $d=7$ it reads 
\begin{align}
\label{eq:7d-int-last}
\frac{3}{8\pi^3}
\int^{\infty}_{-\infty}\,\mathrm{d}u\,\chi(u)
\int^{\infty}_{0}\,\mathrm{d}s\,\chi(u-s)
\left(
\frac{\sin{(Es)}}{\sqrt{-\left[{(\Delta\x)}^2\right]^5}}-\frac{E}{s^4}+\frac{E\left(4E^2+5\ddot{\x}^2\right)}{24s^2}
-\frac{5E\,\ddot{\x}\cdot\x^{(3)}}{24s}
\right)
\ . 
\end{align}

Next, observe that our formulas 
\eqref{eq:6d:detresp}, 
\eqref{eq:5d:detresp:ssform}, 
and 
\eqref{eq:4dFraw}
contain terms in which the subtracted negative powers of $s$ are combined with similar powers of $s$ multiplied by quadratic combinations of $\chi$ and its derivatives evaluated at $u$ rather than at~$u-s$. All the negative powers of $s$ that appear in \eqref{eq:7d-int-last} have already appeared in this fashion in~\eqref{eq:6d:detresp}, and comparison of the coefficients shows that the corresponding terms for $d=7$ read 
\begin{align}
&-\frac{E^3}{16\pi^3}\int^{\infty}_{0}\,\frac{\mathrm{d}s}{s^2}\int^{\infty}_{-\infty}\,\mathrm{d}u\, \chi(u)\left[\chi(u-s)-\chi(u)\right]
\notag
\\[1ex]
&+\frac{3E}{8\pi^3}\int^{\infty}_{0}\,\frac{\mathrm{d}s}{s^4}\int^{\infty}_{-\infty}\,\mathrm{d}u\, \chi(u)
\bigl[
\chi(u-s)-\chi(u)
-\tfrac12 s^2\ddot{\chi}(u)
\bigr]
\notag
\\[1ex]
&-\frac{5}{64\pi^3}\int^{\infty}_{0}\,\frac{\mathrm{d}s}{s^2}
\int^{\infty}_{-\infty}\,\mathrm{d}u\,\chi(u)
\Bigl\{
\left[\chi(u-s)-\chi(u)\right] \ddot{\x}^2
-s \chi(u-s) \, \ddot{\x}\cdot\x^{(3)}
\Bigr\}
\ . 
\label{eq:7d-int-middle}
\end{align} 

The remaining term in 
\eqref{eq:6d:detresp}, 
\eqref{eq:3d:detrespfinal}, 
\eqref{eq:5d:detresp:ssform}, 
and 
\eqref{eq:4dFraw}
is a single integral involving derivatives of~$\x$. 
We are not aware of pattern arguments 
that might fix this term fully for general $d$, 
but we note that if this term for $d=7$ contains the piece 
\begin{align}
\frac{1}{2048\pi^2}\int^{\infty}_{-\infty}\,\mathrm{d}u
\, 
\chi^2(u) 
\left(4E^2+\ddot{\x}^2\right) 
\left(4E^2+9\ddot{\x}^2\right)
\ , 
\label{eq:7d-int-first-part}
\end{align}
then the transition rate computed from 
\eqref{eq:7d-int-last}, 
\eqref{eq:7d-int-middle}
and  
\eqref{eq:7d-int-first-part}
for a uniformly linearly accelerated trajectory agrees with that 
found in~\cite{takagi}. We further note that the power of $E$ in the single integral term in \eqref{eq:6d:detresp}, 
\eqref{eq:3d:detrespfinal}, 
\eqref{eq:5d:detresp:ssform}, 
and 
\eqref{eq:4dFraw}
fits the empirical formula 
\begin{equation}
\frac{\Gamma(d/2-1)}{(d-3)!} \frac{(-E)^{(d-3)}}{4\pi^{(d/2-1)}}
\ , 
\end{equation} 
and so does the highest power of $E$ in~\eqref{eq:7d-int-first-part}. 

We anticipate that the $d=7$ response function contains terms in addition to 
\eqref{eq:7d-int-last}, 
\eqref{eq:7d-int-middle}
and  
\eqref{eq:7d-int-first-part}; 
in particular, the pattern from $d\le6$ suggests that there should be a term proportional to $\delta^{-3}$ as $\delta\to0$, perhaps involving $\int_{-\infty}^{\infty}\mathrm{d}u \, \ddot\chi^2(u)$. 
However, if the only terms contributing to the 
transition rate are \eqref{eq:7d-int-last}, 
\eqref{eq:7d-int-middle}
and 
\eqref{eq:7d-int-first-part}, 
then a comparison with the $d=6$ case shows that the 
transition rate takes the form 
\begin{align}
\dot{\mathcal{F}}_{\tau}\left(E\right)
&=
\frac{5E \, \ddot{\x}(\tau) \cdot \x^{(3)}(\tau)}{64\pi^3}\left(\ln{\left(\frac{\Delta\tau}{\delta}\right) +C^{'}_{+}}\right)
+\frac{\bigl(4E^2+9\ddot{\x}^2(\tau)\bigr)
\bigl(4E^2+\ddot{\x}^2(\tau)\bigr)}{2048\pi^2}
\notag
\\[1ex]
&\hspace{2ex}
+\frac{3}{8\pi^3}\int^{\Delta\tau}_{0}\,\mathrm{d}s 
\left(
\frac{\sin{(Es)}}{\sqrt{-\left[{(\Delta\x)}^2\right]^5}}-\frac{E}{s^4}+\frac{E\bigl(4E^2+5\ddot{\x}^2(\tau)\bigr)}{24s^2}
-\frac{5E\,\ddot{\x}(\tau)\cdot\x^{(3)}(\tau)}{24s}
\right)
\notag
\\[1ex]
&\hspace{2ex}
+\frac{E\bigl(4E^2+5\ddot{\x}^2(\tau)\bigr)}{64\pi^3\Delta\tau}
-\frac{E}{8\pi^3\Delta\tau^3}
+O\left(\delta\ln{\left(\frac{\Delta\tau}{\delta}\right)}\right)
\ , 
\label{eq:7d:guess}
\end{align}
where $C^{'}_{+}$ is again 
given by~\eqref{eq:Cprimeplus}. 
While we must leave \eqref{eq:7d:guess} 
to the status of a conjecture, we note that it shares the 
logarithmic divergence of of the 
$d=6$ transition rate \eqref{eq:ss:6d:transrate} and the 
divergent term is again proportional to $\ddot{\x} \cdot \x^{(3)}$.

\section{Application: Schwarzschild embedded in $d=6$ Minkowski}
\label{sec:GEMS}

The GEMS method
\cite{Deser:1997ri,Deser:1998bb,Deser:1998xb,Santos:2004ws} aims to model detector
response in four-dimensional spacetime by an embedding into a
higher-dimensional flat spacetime with an appropriately-chosen quantum
state, typically the Minkowski vacuum. The method has yielded
reasonable results for stationary trajectories in spacetimes of high
symmetry. A~review with references is given
in~\cite{Langlois,Langlois-thesis}.

We wish to discuss the prospects of GEMS modelling in nonstationary
situations in view of our results.

Recall that the $d=4$ Minkowski vacuum response function formula
\eqref{eq:4dFraw} and instantaneous transition rate formula
\eqref{eq:ss:4d:transrate} generalise to an arbitrary Hadamard state
on an arbitrary four-dimensional spacetime as \cite{satz-louko:curved}
\begin{align}
{\mathcal{F}}(E) 
&=
-\frac{E}{4\pi}\int_{-\infty}^{\infty}\mathrm{d}u\,{[\chi(u)]}^2 
\ + \ 
\frac{1}{2\pi^2}\int_0^{\infty}
\frac{\mathrm{d}s}{s^2}\int_{-\infty}^{\infty}\mathrm{d}u\,\chi(u)
\bigl[ \chi(u)-\chi(u-s)\bigr] 
\nonumber 
\\[1ex]
&\hspace{3ex}
+ 
2
\int_{-\infty}^{\infty}\mathrm{d}u\,\chi(u)
\int_0^{\infty}\mathrm{d}s\,\chi(u-s) 
\,\mathrm{Re} \left(
\mathrm{e}^{-iE s}\, W_0(u,u-s)+\frac{1}{4\pi^2s^2}
\right)
\ , 
\label{eq:4dcurvedprobability}
\\[1ex]
\label{eq:4dcurvedrate}
\dot{\mathcal{F}}_{\tau}\left(E\right)
&=
-\frac{E}{4\pi}
+2\int_0^{\Delta\tau}\mathrm{d}s
\, \mathrm{Re}
\left( \mathrm{e}^{-iE s}W_0(\tau,\tau-s)+\frac{1}{4\pi^2s^2}\right)
\ \ +\frac{1}{2\pi^2 \Delta \tau} 
\ , 
\end{align}
where $W_0$ is the pointwise 
$i\epsilon\to0$ limit of the Wightman function. 
The divergence structure at $\delta\to0$ is exactly 
as in Minkowski vacuum: 
the response function 
\eqref{eq:4dcurvedprobability} diverges logarithmically but the transition 
rate has the finite limit given by~\eqref{eq:4dcurvedrate}. 

As a concrete example, consider a detector in the extended
Schwarzschild spacetime, globally embedded in $d=6$ Minkowski space as
in \cite{fronsdal} (for further discussion see~\cite{ferraris}).  For
static trajectories in exterior Schwarzschild, GEMS modelling with
$d=6$ Minkowski vacuum predicts a thermal response in the local
Hawking temperature~\cite{Deser:1998xb}. One might hence anticipate
this modelling to extend to more general detector trajectories in the
Hartle-Hawking-Israel vacuum \cite{Hartle:1976tp,Israel:1976ur}.

Now, while the genuine $d=4$ sharp switching transition rate
\eqref{eq:4dcurvedrate} is finite for arbitrary trajectories in the
Hartle-Hawking-Israel vacuum, the $d=6$ Minkowski vacuum transition
rate \eqref{eq:ss:6d:transrate} diverges in the sharp switching limit
unless the $d=6$ scalar proper acceleration is constant. There are
trajectories of constant $d=6$ scalar proper acceleration through
every point in the extended Schwarzschild spacetime, and these
trajectories include all the stationary trajectories, that is, the
exterior region circular trajectories that have constant (in general
non-inertial) angular velocity. However, we have verified by a direct
calculation that the only timelike Schwarzschild geodesics of constant
$d=6$ scalar acceleration are the exterior circular geodesics.  This
suggests that the GEMS method may not provide a viable model for
detectors on generic geodesics in Schwarzschild.

\section{Summary and discussion}
\label{sec:summary}

We have investigated the response of an arbitrarily-accelerated
Unruh-DeWitt detector coupled to a massless scalar field in Minkowski
spacetimes of dimension up to six, assuming the field to be initially
in the Minkowski vacuum and the detector to be switched on and off
smoothly. Following the four-dimensional analysis
of~\cite{satz:smooth}, we first expressed the response function as a
manifestly regular and Lorentz-covariant integral formula from which
the $i\epsilon$ regulator of the Wightman function has been
eliminated. We then examined the sharp switching limit of the response
function and of the transition rate, defined as the derivative of the
response function with respect to the total detection time.

In four dimensions it was shown in \cite{satz:smooth} that the
response function diverges in the sharp switching limit
as~$\ln\delta$, where $\delta$ is the duration of the switch-on and
switch-off periods, while the transition rate remains finite in this
limit. In two dimensions it is immediate from the merely logarithmic
singularity of the Wightman function that both the response function
and transition rate remain finite in the sharp switching limit. In
three dimensions we found that both the response function and the
transition rate remain finite in the sharp switching limit. In five
dimensions the response function diverges in the sharp switching limit
as~$1/\delta$, but the transition rate remains finite.

In six dimensions a qualitatively new phenomenon emerges. 
The response function diverges in the sharp switching 
limit as~$1/\delta^2$, but now also the transition rate 
contains a divergent term, proportional to 
$\ddot{\x}\cdot\x^{(3)} \ln\delta$; 
further, even if $\delta$ is kept nonvanishing, 
the coefficient of $\ddot{\x}\cdot\x^{(3)}$
depends on the details of the switch-off profile. 
The transition rate hence diverges for generic trajectories, 
although it remains finite for trajectories on which 
the scalar proper acceleration is constant, 
including all stationary trajectories. 

To summarise, the results show that the instantanenous transition rate
of the Unruh-DeWitt detector is well defined in dimensions up to
five. In dimension six the class of trajectories for which the
transition rate is well defined includes all stationary trajectories,
and among them in particular the Rindler trajectories of uniform
linear acceleration for which the Unruh effect arises. For generic
trajectories in six dimensions the transition rate is however
divergent.

We have not pushed the computations beyond six dimensions, but we note
that the divergences in the sharp switching limit arise from the
singularity of the Wightman function, and this singularity grows in
strength as the spacetime dimension increases. We therefore anticipate
that the instantaneous transition rate will continue to be divergent
for generic trajectories in all dimensions above five.  We also
anticipate that this property holds not just for Minkowski vacuum in
Minkowski spacetime but for an arbitrary Hadamard state in an
arbitrary spacetime. Finally, our explicit formulas for the response
function and the transition rate in three dimensions are generalised
to an arbitrary Hadamard state in an arbitrary spacetime
in~\cite{Hodgkinson:2012mr}, using the techniques that allowed such a
generalisation in four dimensions~\cite{satz-louko:curved}, and we
anticipate a similar generalisation to be possible also in five and
six dimensions.

While the divergence of the transition rate beyond five dimensions may
be mathematically surprising, we emphasise that this does not seem to
indicate any physical pathology in the system. It just tells that a
\emph{pointlike\/} detector is too singular to have a well-defined a
transition rate.  There appears to be no difficulty with defining the
transition rate of a spatially smeared
detector~\cite{Langlois,Langlois-thesis}: the difficulty surfaces only
in the zero spatial size limit, as we discussed in
section~\ref{sec:consistencychecks}.

As an application of our results in six dimensions, we investigated
modelling a particle detector in four-dimensional Schwarzschild by the
GEMS technique, embedding Schwarzschild in six-dimensional Minkowski
and setting the six-dimensional quantum field initially in Minkowski
vacuum.  This modelling yields a well-defined transition rate for all
stationary trajectories in Schwarzschild.  However, we found that the
only Schwarzschild geodesics that are mapped to trajectories of
constant scalar proper acceleration in six-dimensional Minkowski are
the circular geodesics. The modelling does hence not yield
well-defined transition rates for any noncircular geodesics in
Schwarzschild.  This, together with the anticipated extensions of our
results beyond six dimensions, suggests that the GEMS technique may
have limited validity for nonstationary trajectories whenever the
embedding spacetime has dimension higher than five.

We have assumed the detector's trajectory 
to be smooth on the support of the switching function.  
The first role of this assumption is that it ensures 
the pull-back of the Wightman function to the 
detector's worldline to be a  well-defined distribution 
that can be represented by a family of functions with 
a standard $i\epsilon$ regulator. The second role 
is that it allows Taylor expansions in 
the proper time to be used to arbitrarily high order. 
If the $i\epsilon$ representation 
of the pull-back of the Wightman function  
can be justified for a trajectory of lower 
differentiability class, then the practical constraint on the differentiability class for our methods to 
remain applicable is that the Taylor expansions in proper time, 
such as those in 
\eqref{eq:6d:odd:premult}
and 
\eqref{eq:6d:even:premult}, still hold to the required order. 

On a practical note, 
we remark that we have made essential 
use of algebraic computing with the Taylor 
series expansions, 
and to a lesser extent also with
the numerous elementary integrals. 
As a check against errors, 
we have performed a number of these 
computations independently on 
both Maple and Mathematica. 

We have throughout the paper worked in linear perturbation theory in
the coupling between the field and the detector.  Going beyond
perturbation theory would raise a number of new questions.  Would a
smooth switch-on and switch-off regularise the response of a pointlike
detector also in a fully nonperturbative treatment? Could a transition
rate for the detector be recovered in a sharp switching limit, at
least in sufficiently low spacetime dimensions? How would the results
compare with those obtained in a nonperturbative treatment with sharp
switching but with a time sampling
regulator~\cite{Lin:2006jw,Ostapchuk:2011ud}?

\section*{Acknowledgements}

We thank Michael White for discussions and for 
collaboration in the early stages of the work~\cite{white-dissertation}, 
including the discovery of the choice 
$\eta=\epsilon^{1/4}$ for estimating~\eqref{eq:detres6d}. 
We also thank 
Bernard Kay, 
Shih-Yuin Lin, 
Robb Mann 
and 
Alejandro Satz 
for useful discussions. 
J.~L. was supported in part by 
the Science and Technology Facilities Council.

\end{document}